\documentclass[twocol]{arxiv_ams}

\usepackage{pdfpages}

\usepackage{amsmath,amsfonts,amssymb,bm}
\usepackage{mathptmx}
\usepackage{newtxtext}
\usepackage{newtxmath}
\usepackage{sidecap}

\usepackage{comment}
\title{Reanalyses and a high-resolution model fail to capture the `high tail' of CAPE distributions \\ \textcolor{red}{Submitted to Journal of Climate, under review} }

\authors{ZIWEI WANG}
\affiliation{Department of the Geophysical Sciences, University of Chicago, 
    Chicago, Illinois \\
    Center for Robust Decision-making on Climate and Energy Policy (RDCEP), University of Chicago, Chicago, Illinois}

\extraauthor{JAMES A.\ FRANKE}
\extraaffil{Department of the Geophysical Sciences, University of Chicago,
Chicago, Illinois  \\
Center for Robust Decision-making on Climate and Energy Policy (RDCEP), University of Chicago, Chicago, Illinois}

\extraauthor{ZHENQI LUO }
\extraaffil{College of Natural Resources, Faculty of Geographical Science, Beijing Normal University, Beijing, China}

\extraauthor{ELISABETH J.\ MOYER \correspondingauthor{Elisabeth J.\ Moyer, moyer@uchicago.edu}}
\extraaffil{Department of the Geophysical Sciences, University of Chicago, 
Chicago, Illinois \\  
Center for Robust Decision-making on Climate and Energy Policy (RDCEP), University of Chicago, Chicago, Illinois}

\abstract{
Convective available potential energy (CAPE) is of strong interest in climate modeling because of its role in both severe weather and in model construction. Extreme levels of CAPE ($>$ 2000 J/kg) are associated with high-impact weather events, and CAPE is widely used in convective parametrizations to help determine the strength and timing of convection. However, to date no study has systematically evaluated CAPE biases in models in a climatological context, in an assessment large enough to characterize the high tail of the CAPE distribution. This work compares CAPE distributions in over 200,000 summertime proximity soundings from four sources: the observational radiosonde network (IGRA), 0.125 degree reanalysis (ERA-Interim and ERA5), and a 4 km convection-permitting regional WRF simulation driven by ERA-Interim. Both reanalyses and model consistently show too-narrow distributions of CAPE, with the high tail ($>$ 95th percentile) systematically biased low by up to 10\% in surface-based CAPE and 20\% at the most unstable layer.
This ``missing tail'' corresponds to the most impacts-relevant conditions. CAPE bias in all datasets is driven by bias in surface temperature and humidity: reanalyses and model undersample observed cases of extreme heat and moisture. These results suggest that reducing inaccuracies in land surface and boundary layer models is critical for accurately reproducing CAPE.
}

\begin{document}
\begin{figure}
\includegraphics[width=\textwidth]{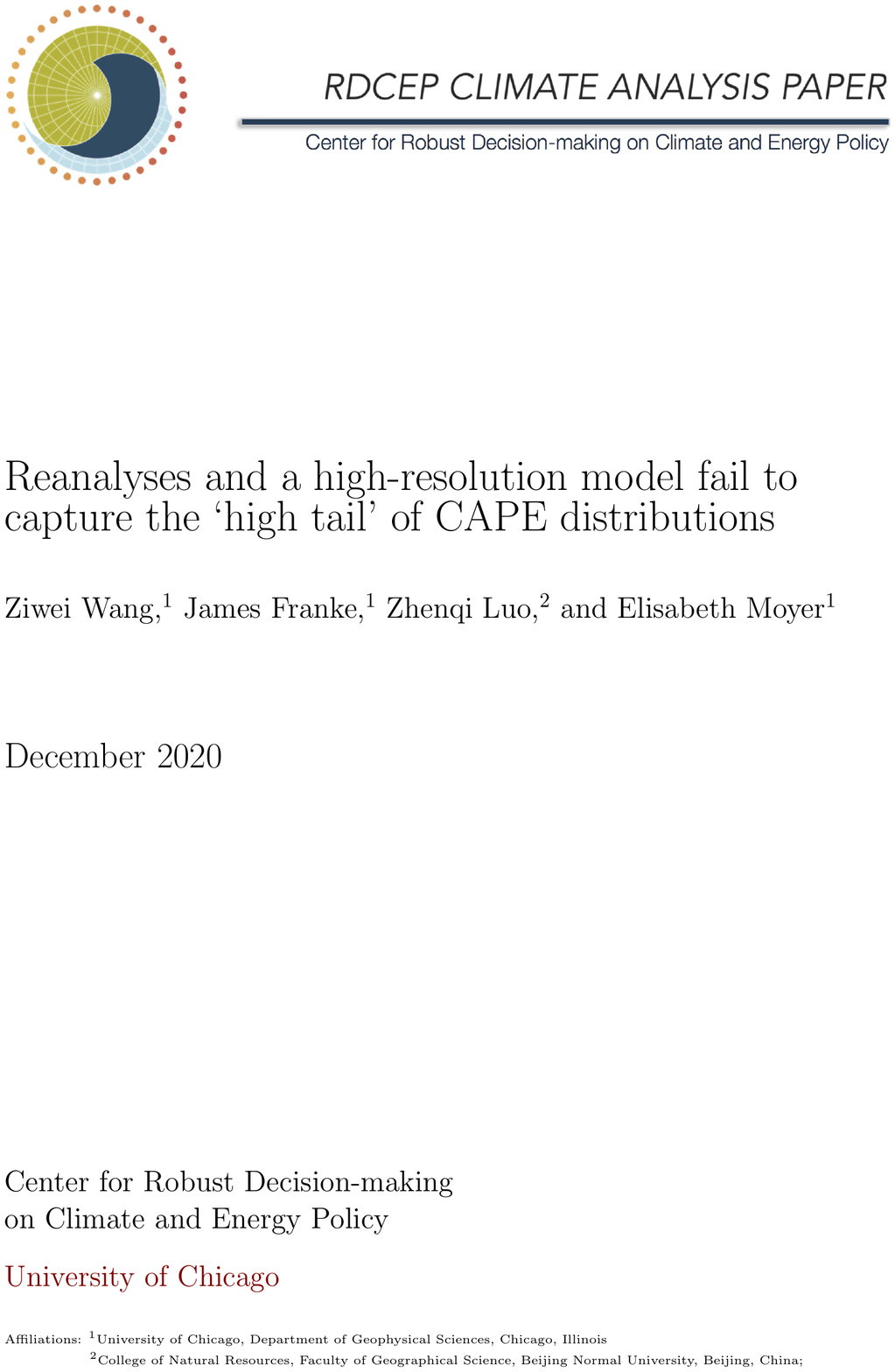}
\end{figure}
\clearpage
\begin{figure}
\includegraphics[width=\textwidth]{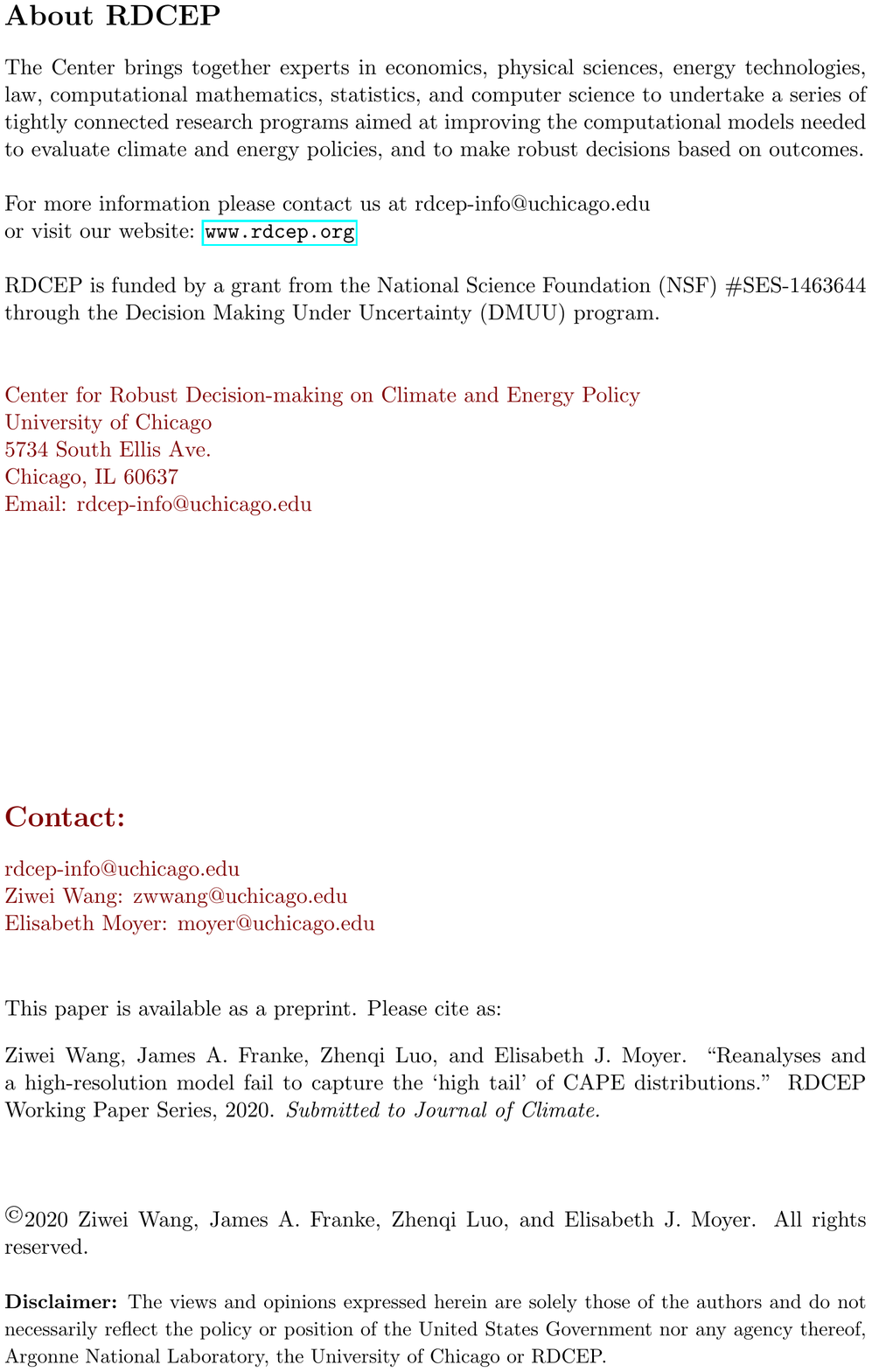}
\end{figure}
\newpage

%% Necessary!
\maketitle

%%%%%%%%%%%%%%%%%%%%%%%%%%%%%%%%%%%%%%%%%%%%%%%%%%%%%%%%%%%%%%%%%%%%%
% SIGNIFICANCE STATEMENT/CAPSULE SUMMARY
%%%%%%%%%%%%%%%%%%%%%%%%%%%%%%%%%%%%%%%%%%%%%%%%%%%%%%%%%%%%%%%%%%%%%
%
% If you are including an optional significance statement for a journal article (as 1/1/2020, significance statements are only applicable for WCAS or WAF) 
% or a required capsule summary for BAMS 
% (see www.ametsoc.org/ams/index.cfm/publications/authors/journal-and-bams-authors/formatting-and-manuscript-components for details), 
% please apply the necessary command as shown below:
%
% \statement
% Significance statement here.
%
% \capsule
% Capsule summary here.

%%%%%%%%%%%%%%%%%%%%%%%%%%%%%%%%%%%%%%%%%%%%%%%%%%%%%%%%%%%%%%%%%%%%%
% MAIN BODY OF PAPER
%%%%%%%%%%%%%%%%%%%%%%%%%%%%%%%%%%%%%%%%%%%%%%%%%%%%%%%%%%%%%%%%%%%%%
%

%% In all cases, if there is only one entry of this type within
%% the higher level heading, use the star form: 
%%
% \section{Section title}
% \subsection*{subsection}
% text...
% \section{Section title}

%vs

% \section{Section title}
% \subsection{subsection one}
% text...
% \subsection{subsection two}
% \section{Section title}

%%%
% \section{First primary heading}

% \subsection{First secondary heading}

% \subsubsection{First tertiary heading}

% \paragraph{First quaternary heading}

\section{Introduction}

% This paragraph: Extreme weather and CAPE. 
Convective Available Potential Energy (CAPE) is an integral quantity of buoyancy in the convective layer \citep{moncrieff_dynamics_1976}, and is considered as a key parameter in convection initiation and development. Closely linked to updraft strength and storm intensity, CAPE provides a way to understand the potential threat of some high-impact weather events such as thunderstorms, hail, and tornadoes. \citet{brooks_spatial_2003} proposed a combination of CAPE and bulk wind shear as a metric for severe weather in reanalyses, with a 2000 J/kg as a threshold value for extreme events, and multiple subsequent studies confirm this relationship in models and observations.  Studies relating high CAPE values to extreme precipitation or intense storms in observations include \citet{groenemeijer_sounding-derived_2007}, \citet{lepore_temperature_2015}, \citet{dong_precipitable_2019}, and many others.  In models, \citet{paquin_change_2014}, for example, show that the number of extreme precipitation events in general circulation models (GCMs) grows with the covariate between CAPE and wind shear.

% CAPE is important in convective parametrization
CAPE is also used as a key parameter in convective schemes in GCMs to determine convective mass flux \citep{zhang_sensitivity_1995,yano_phenomenology_2013,baba_spectral_2019}. In CAPE--closure (or CR--closure) schemes, modelers rely on CAPE to trigger convection and to determine the total vertical mass flux.  While the timing of convection onset in most schemes does not depend on exact CAPE values, and is driven instead by a range of conditions including dynamics and thermodynamics fields \citep{yang_simulated_2018}, the magnitude of vertical mass flux is directly affected by an inaccurate representation of CAPE \citep{lee_role_2008, cortes-hernandez_evaluating_2016}. In some recently developed new schemes intended to more realistically reproduce the diurnal cycle, convective triggering is directly dependent on CAPE generation rate (dCAPE) \citep{xie_impact_2000,wang_impacts_2015}. These schemes have been shown to improve model performance for precipitation diurnal peak time \citep{song_improving_2017,xie_improved_2019}, but introduce additional sensitivity to CAPE biases.  

% Direct measurement for CAPE calculation: Radiosondes - but most people use reanalysis
CAPE is derived from vertical profiles of temperature, pressure, and humidity, which are measured in-situ only from a sparse network of specialized weather stations. Radiosondes measure atmospheric profiles from weather balloons released twice a day from $\sim$ 1000 stations globally (77 in the contiguous U.S. still in service). Because radiosonde measurements are both spatially and temporally sparse, researchers linking measured CAPE to severe weather events have used ``proximity soundings'', estimating the severity of extreme weather events based on  soundings taken within a range of $\sim$200 km (e.g. \citealt{brooks_environments_1994,rasmussen_baseline_1998,Brooks200216}). More recent studies of CAPE and severe weather use not soundings but reanalyses that assimilate in-situ and remote observations in global models to provide information at higher resolution \citep{brooks_spatial_2003, lepore_temperature_2015, dong_precipitable_2019}. Global gridded reanalyses also allow ready construction of climatologies: for example, \citet{riemann-campe_global_2009} use the ERA40 reanalysis to construct a 40-year climatology of CAPE, showing that largest values and variability are found over tropical land (mean $\sim$2000 J/kg), with a stronger dependence on specific humidity than temperature. 

% and some people just look at models
To diagnose potential changes in CAPE under future higher CO$_2$ conditions, studies must rely on numerical simulations. To get a basic sense of model performance under current climate, \citet{chen_changes_2020} validates the ability of GCM (CCSM4) to realistically capture spatial pattern of CAPE and CIN in reanalysis, but also identifies discrepancy even for the mean values up to 500 J/kg.
With the growth of computational resources, the horizontal resolution of models used for this purpose have increased. For example, \citet{trapp_transient_2009} and \citet{diffenbaugh_robust_2013} examine changes in CAPE and wind shear in GCM projections ($\sim$100 km) and infer a likely future increase in the number of days with severe weather events. \citet{singh_increasing_2017} use both GCMs and super-parametrized GCMs (20 km) to study changes in the the 95th percentile of CAPE in the tropics and subtropics during heavy precipitation, and find a 6--14\% increase per K regional temperature increase. [Note that CAPE values during heavy precipitation are low, e.g. \citet{adam_cape_2009}; the 95th percentile in observations in \citet{singh_increasing_2017} is under 2000 J/kg.] \citet{rasmussen_changes_2017} examine changes in CAPE and convective inhibition (CIN) in a 4 km dynamically downscaled simulation of North America in a pseudo global warming scenario (driven by reanalysis or by reanalysis with an applied offset in climate variables). They find that both CAPE and CIN generally increase under warmer conditions, and infer a future intensification of convective strength. Such cloud resolving models, with their improvement in convective dynamics, have been assumed to help improve the representation of CAPE. 

% CAPE bias?
Given the extent of scientific use of reanalyses and model simulations, it is valuable to ask how well these products reproduce realistic CAPE values. Several studies assess CAPE bias in reanalysis or forecast models versus radiosonde observations, but all use restricted samples of soundings near severe weather events, and study results are inconsistent. \citet{thompson_close_2003} evaluate surface-based CAPE (SBCAPE) from the Rapid Update Cycle (RUC-2) weather prediction system 0-hour analysis against radiosondes sampled near supercells (149 soundings from 1999--2001, in the U.S.\ Central and Southern Plains) and find a low bias of $\sim$16\% (mean bias of about -400 J/kg in mean conditions of $\sim$2500 J/kg). \citet{coniglio_verification_2012} compare SBCAPE in the RUC 0-hour analysis with a different sample  of soundings near supercell thunderstorms (582 soundings during the VORTEX2 campaign in 2009--2010, also in the Central and Southern Plains) and find a small high bias ($\sim$150 J/kg) with large spread. \citet{allen_climatology_2014} compare mixed-layer CAPE (MLCAPE) in the reanalysis product ERA-Interim (ERAI) and in the Australian MesoLAPS (Mesoscale Limited Area Prediction System) weather model with radiosonde soundings near thunderstorm events (3697 and 4988 soundings, respectively, from 2003--2010, from 16 stations in Australia) and find slight high biases of 6 and 74 J/kg in conditions of 234 and 255 J/kg mean non-zero MLCAPE.

Many authors attribute errors in CAPE to incorrect temperature and humidity at the surface or boundary layer. Several studies have explicitly tested this attribution by replacing surface values in models and data products with observed ones and noting the improved match to radiosonde SBCAPE.  \citet{coniglio_verification_2012} replaces  surface values in RUC with those from the operational surface objective analysis system (SFCOA), and finds a reduction in bias in 1-hour forecasts. \citet{gartzke_comparison_2017} compare 10 years of SBCAPE from a single station, the Southern Great Plains Atmospheric Radiation Measurement (ARM) site, and show that replacing surface values largely corrects CAPE values in ERAI reanalysis and values derived from the AIRS satellite. Similarly, in a very small sample (2 individual case studies), \citet{bloch_near-real-time_2019} finds that replacing surface values of humidity and temperature corrects a low bias in SBCAPE in a satellite-derived product. 

To date, no validation study has systematically evaluated CAPE bias and errors in a climatological context, with a large enough scale to allow evaluation of the high tail of the CAPE distribution. \citet{gensini_severe-thunderstorm_2013} do consider a wide selection of soundings and conditions, comparing NARR (the North American Regional Reanalysis) to all radiosondes over 11 years from 21 stations in the Eastern U.S. ($>$100,000 soundings with nonzero SBCAPE from 2000--2011), but do not assess either mean bias or distributional differences. (They do find considerable spread in SBCAPE errors, with RMSE $\sim$1400 J/kg.)  In cloud-resolving models, the assumption that improved resolution also improves the representation of CAPE has not been explicitly tested. This work seeks to address the need for a systematic validation of CAPE in reanalyses and high-resolution simulation results by comparing these values against a large radiosonde dataset, using 12 years of observations (2001--2012) from 80 stations over the contiguous U.S. 

\section{Data Description} 
This study compares four datasets that allow calculation of CAPE over the contiguous United States from January 2001 to December 2012: radiosonde observations from the Integrated Global Radiosonde Archive (IGRA) version 2; the reanalysis products ERA-Interim (ERAI) and ERA5; and simulation output from the Weather Research and Forecasting model (WRF) at convection-permitting resolution, forced by ERAI \citep{Rasmussen_Liu}. Because our interest is in the high tail of the CAPE distribution, we focus on the summer months when convection is most active and CAPE is largest. We define summer as May to August (MJJA), following the convention of many studies (e.g.\ \citealt{sun_evaluation_2016,rasmussen_changes_2017}), though some work on extreme weather uses an earlier definition of April to July to include the late spring peak of convection  (e.g.\, \citealt{trapp_transient_2009}). With this definition, IGRA provides a total of 199,787 summertime radiosonde profiles from U.S.\ stations with continuous records during 2001--2012. For consistency, analyses shown here involve data matched to these profiles, using the nearest output to each radiosonde station and generally synchronized in time, though when evaluating diurnal cycles we also show reanalysis and model output at additional times of day.

\subsection{Radiosonde observations}
IGRA is an archive of quality-controlled atmospheric sounding profiles from weather balloons around the world collected by a standard protocol  \citep{durre_overview_2006,durre_robust_2008}. The archive is operated by the U.S.\ National Oceanic and Atmospheric Administration (NOAA) and profiles in the U.S.\ are collected by NOAA's National Weather Service. In this work we use profiles from all stations in the contiguous United States that report continuous operation through the years 2001 to 2012, a total of 80 out of the 248 stations historically used. All stations have routine balloon launches at 00 and 12 UTC each day, though some soundings are missing (17.4\% of all routine launches during this period). Many stations also include sporadic launches at 06 and 18 UTC; we include these profiles in the dataset considered here, though we generally disaggregate analyses by time of day. Of the complete dataset of 199,787 soundings, 83,668 are from 00 UTC, 106,455 from 12 UTC, and 9,664 from additional times.  Of these profiles, 245 (0.14\%) are excluded by our quality control criteria. (See Methods below.)

Variables acquired from IGRA include pressure, temperature, altitude, and vapor pressure, all of which are standard reported values. We convert vapor pressure to specific humidity and dew point temperature for consistency across all datasets.  Vertical resolution varies by station, but most stations report around 90 levels from surface to 10 hPa pressure. The data are available from \url{https://www.ncdc.noaa.gov/data-access/weather-balloon/integrated-global-radiosonde-archive}.

\subsection{Reanalysis products}
ERAI and ERA5 are both reanalysis products maintained by the European Centre for Medium-Range Weather Forecasts (ECMWF). Both products assimilate observations into global models and are available from 1979 to the present. ERAI has a native horizontal resolution of T255 ($\approx$ 80km) \citep{dee_era-interim_2011}; it is been superseded by ERA5, which has significant improvements in spatial and temporal resolution with a native horizontal resolution of TL639 (0.28125$^\circ$, $\approx$ 31km)  \citep{C3S}. Because our analysis involves matching individual radiosonde stations, we acquire both reanalyses at a finer spatial resolution (0.125$^\circ$) produced by ECMWF with bilinear interpolation for continuous fields. We use output at native model vertical levels, preserving the highest possible vertical resolution for our CAPE calculation: 60 levels for ERAI (L60), and 137 for ERA5 (L137). We download profiles of temperature and specific humidity, and surface pressure; the pressure profile is then derived using surface pressure and scaling factors provided by ECMWF (\url{https://www.ecmwf.int/en/forecasts/documentation-and-support}). 2m temperature and dew point temperature along with surface pressure are appended to the bottom level of profiles. Although ERA5 provides hourly output, we use data at 00, 06, 12 and 18 UTC to match with ERAI. Both products are available at \url{https://www.ecmwf.int/en/}.

% One paragraph to say about the assimilation
Data assimilation is a key component of reanalysis products. Both ERAI and ERA5 assimilate a homogenized version of IGRA radiosonde observations, the Radiosonde Observation Correction using Reanalyses (RAOBCORE) \citep{haimberger_homogenization_2007,haimberger_toward_2008}. Reanalyses and IGRA observations are therefore not fully independent. ERAI uses a bias correction for radiosonde temperature based on RAOBCORE\_T\_1.3, which is further adjusted and implemented to the Continuous Observation Processing Environment (COPE) framework in ERA5 \citep{ECMWF2016}. The assimilation process of ERAI uses the following exclusion criteria for radiosonde data: 1) any radiosonde observation below the model surface, and radiosonde-observed specific humidity in either 2) extreme cold conditions (T $<$ 193 K for RS--90 sondes, T $<$ 213 K for RS--80 sondes, T $<$ 233 K otherwise), or 3) high altitude (p $<$ 100 hPa for RS--80 and RS--90 sondes, p $<$ 300 hPa for all other sonde types) \citep{dee_era-interim_2011}. 

\subsection{High-resolution model simulation}
The high-resolution model output we use is a 4-km resolution dynamically downscaled ``retrospective'' simulation over North America first described by \citet{liu_continental-scale_2017}. The simulation was created as the control run of a pseudo-global-warming experiment, and involves forcing the WRF (Weather Research and Forecasting) 3.4.1 model with ERAI reanalysis. The WRF simulation is run with 4 km grid spacing and 50 vertical levels up to 50 hPa, with parametrization schemes including: Thompson aerosol-aware microphysics \citep{thompson_study_2014}, the Yonsei University (YSU) planetary boundary layer \citep{hong_new_2006}, the rapid radiative transfer model (RRTMG) \citep{iacono_radiative_2008}, and the improved Noah-MP land-surface model \citep{niu_community_2011}. 

The model uses ERAI as initial and boundary conditions, with spectral nudging applied to geopotential, temperature and horizontal wind. Nudging is applied throughout the model domain, at all altitudes above the planetary boundary layer, to remove known issues of summertime high temperature bias over the central U.S.\ \citep{morcrette_introduction_2018}.  Values are nudged at a strength corresponding to an `e-folding' time of 6 hours, using a wavenumber truncation of 3 and 2 in the zonal and meridional directions, respectively. Because the experiment was intended to reproduce observed snow cover over North America, some modifications were made to the land surface model, including representing the heat transport from rainfall caused by the temperature difference between raindrops and land surface, and modifying the snow cover/melt curve to produce more realistic surface snow coverage and reduce wintertime low bias in temperature. 
 
Model output is acquired from the NCAR Research Data Archive ds612.0 \citep{Rasmussen_Liu}. We take pressure, temperature, mixing ratio, height from the CTRL 3D subset, and surface topography, surface pressure, 2m temperature and mixing ratio from the CTRL 2D subset.

\section{Methods}
\subsection{CAPE calculation}
All CAPE values shown in this work are calculated with SHARPpy (the Sounding and Hodograph Analysis and Research Program in Python) version 1.4.0a4, a widely used collection of sounding and hodograph analysis routines designed to provide free and consistent analysis tools for the atmospheric sciences community (\url{https://github.com/sharppy/SHARPpy}, \citealt{blumberg_sharppy:_2017}). SHARPpy is an extension of SHARP, which was first released in 1991 \citep{Hart1991}. CAPE in the SHARPpy package is calculated following the definition of \citet{moncrieff_dynamics_1976} in which temperature is automatically corrected to virtual temperature \citep{doswell_effect_1994}. The required variables are vertical profiles of pressure, temperature, height and dew point temperature. Wind speed and direction are optional and we do not include them. The package can produce the CAPE of parcels either at surface level (SBCAPE), at the ``most unstable'' level (MUCAPE) or using the averaged properties of ``mixed layer'' (MLCAPE). SHARPpy is the most commonly used package in the CAPE literature (e.g.\ \citealt{gartzke_comparison_2017, king_north_2019}), which provides a comprehensive list of convective indices as output.

We evaluate CAPE for all summertime profiles corresponding to radiosonde soundings other than those with the following exclusion criteria: 
1) no surface level measurements (0.004\% of soundings), 
2) excessive discrepancy of relative humidity between the surface level and one level above, i.e.\ $RH_{sfc} - RH_{lev1} > 65\%$
(0.012\% of soundings), or 3) less than 10 vertical levels of observations (0.12\% of soundings).
In some cases radiosonde profiles involve missing values in the height variable, even though temperature, pressure, and humidity are reported. In these cases we interpolate height based on pressure  using the  SHARPpy ``INTERP'' function.

\subsection{Testing sensitivity to vertical interpolation}
In the analysis here we interpolate only where data are missing in radiosonde profiles. The number of vertical levels used is therefore inconsistent across datasets. Other authors of CAPE comparison studies have chosen to interpolate to produce consistent vertical sampling, e.g. \citet{gartzke_comparison_2017} who use 202 fixed levels (2 and 30 meters, followed by 75 m spacing from 75 m to 15 km). We test the robustness of derived CAPE to this interpolation; results are shown in Table \ref{tab:interpolation}. While interpolation can matter for some individual profiles (the maximum change is 25\%, 500 J/kg when mean radiosonde CAPE is 2000 J/kg), on average interpolation results in only a trivial ($<$ 1\%) departure from values calculated at native vertical resolution. (See \citealt{coniglio_verification_2012} for similar conclusions.)

\begin{table}[h!]
\begin{center}
\resizebox{\linewidth}{!}{\begin{tabular}{ c | c | c | c| c  } 
\hline
 & IGRA & ERAI & ERA5 & WRF\\ 
\hline
Raw CAPE & 303.2 & 312.0 & 328.4 & 316.5 \\ 
\hline
Interp.\ CAPE & 302.8 &  312.9 & 327.8 & 316.8 \\
\hline
Fractional Diff & -0.13\% & 0.29\% & -0.18\% & 0.09\% \\
\hline
\end{tabular}}  \\
\caption{Test of the sensitivity of SBCAPE to vertical interpolation to higher resolution. Table compares mean CAPE values for year-2012 profiles in our dataset calculated at both native resolution and at 202 fixed levels. Native vertical levels in ERAI, ERA5, and WRF are  61, 138, and 51, respectively. In IGRA, soundings typically have $\sim$80 levels but can vary. All calculations are made with SHARPpy using the ``INTERP'' function for interpolation. Interpolation alters mean values by no more than 0.3\%.
}
\label{tab:interpolation}
\end{center}
\end{table}

\subsection{CAPE definitions} 
CAPE is the potential buoyancy of an parcel lifted to its level of free convection, but the parcel considered may be either one located at the surface (SBCAPE) or at the most unstable vertical level (MUCAPE). Alternatively, CAPE can also be calculated as the mean value for parcels in the entire mixed layer, the lowest 100 hPa of the atmosphere (MLCAPE). All are standard outputs of SHARPpy, and the appropriate choice differs according to the scientific question addressed. We use SBCAPE in most of this work for consistency with prior analyses, but also compare with alternate definitions. Most prior CAPE comparison studies have involved SBCABE (e.g.\ \citealt{gensini_severe-thunderstorm_2013,gartzke_comparison_2017,singh_increasing_2017}), and at least some CR-closure convective parametrizations use SBCAPE (e.g.\ \citealt{xie_impact_2000,wang_impacts_2015}). However, some authors argue that MUCAPE or MLCAPE are more appropriate for characterizing upper layer instability \citep{bunkers_importance_nodate,brooks_climatological_2007}, and \citet{rasmussen_changes_2017} use MLCAPE in their study of the high-resolution WRF output.  

\begin{figure}[t]
 \includegraphics[width=\linewidth]{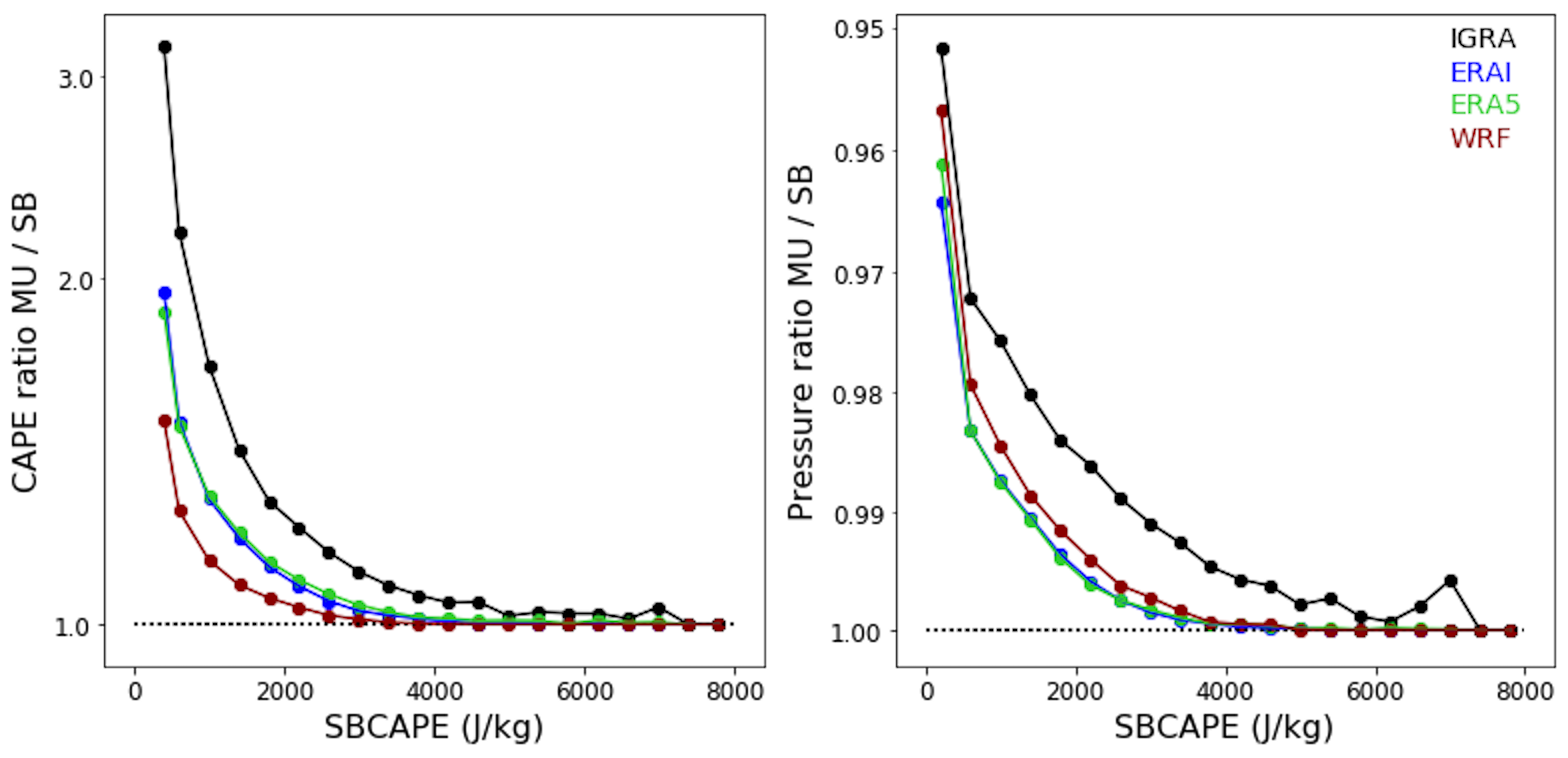}
 \caption{Comparison of SBCAPE and MUCAPE for all datasets, using all soundings considered. Data is binned by SBCAPE value, and we exclude values under 200 J/kg.  (Left) Mean ratio of MUCAPE over SBCAPE, and (right) mean of ratio of the most unstable pressure level over surface pressure. Note that y axes are log scale. For both CAPE and pressure level, the ratio approaches 1.0 as CAPE increases: in higher CAPE conditions, the most unstable level is closer to the surface.}
 \label{fig:levelratio}
\end{figure}

To understand the implications of the different definitions, we compare surface-based CAPE with that of the most unstable layer, MUCAPE, the maximum possible value for each profile (Figure \ref{fig:levelratio}).  Because our focus is on incidences of very high CAPE, we are especially interested in whether different CAPE definitions lead to different understanding of the high tail. Figure \ref{fig:levelratio} shows that the higher the CAPE value, the closer to the surface the most unstable layer becomes, and the more similar SBCAPE and MUCAPE. All datasets show a similar pattern. In conditions conducive to extreme weather ($>$ 4000 J/kg),  SBCAPE and MUCAPE become essentially identical in reanalysis and model output. In radiosonde observations, the distinction between SBCAPE and MUCAPE is greater in all conditions, and the most unstable layer occurs at lower pressures (higher above the surface). 
In conditions with SBCAPE $\sim$1000 J/kg, for example, the average most unstable parcel in radiosonde soundings lies $\sim$30 hPa above the surface, but only $\sim$10 hPa in reanalyses and model. Model and reanalysis biases in MUCAPE will therefore be consistently more negative than those in SBCAPE, though differences are 5\% or less when SBCAPE exceeds 4000 J/kg. 

\section{Results -- biases in CAPE distributions}

\subsection{CAPE distributions across datasets} 
Comparison of the distribution of CAPE in the datasets considered shows immediately that reanalyses and model output underpredict incidences of very high CAPE. Table \ref{tab:distributionSB}  shows the breakdown of SBCAPE above or below threshold values, and Table \ref{tab:distributionMU} the same for MUCAPE. In all datasets, CAPE distributions are zero-peaked, i.e.\ a large fraction ($\sim$40\%) of cases involve zero CAPE, even in the highly convective summertime. The frequency of zero CAPE is broadly similar across datasets,  but in reanalysis and model, incidences of extreme CAPE drop off sharply, with values above 4000 J/kg  substantially underpredicted in both definitions. For SBCAPE, reanalysis and model produce 20--30\% fewer incidences of values $>$ 4000 J/kg. For MUCAPE, the underprediction is even more severe, with 60--70\% of all incidences missed.

%\clearpage
\begin{figure}[t]
 \includegraphics[width=\linewidth,trim={0.5cm 1cm 1cm 1cm},clip]{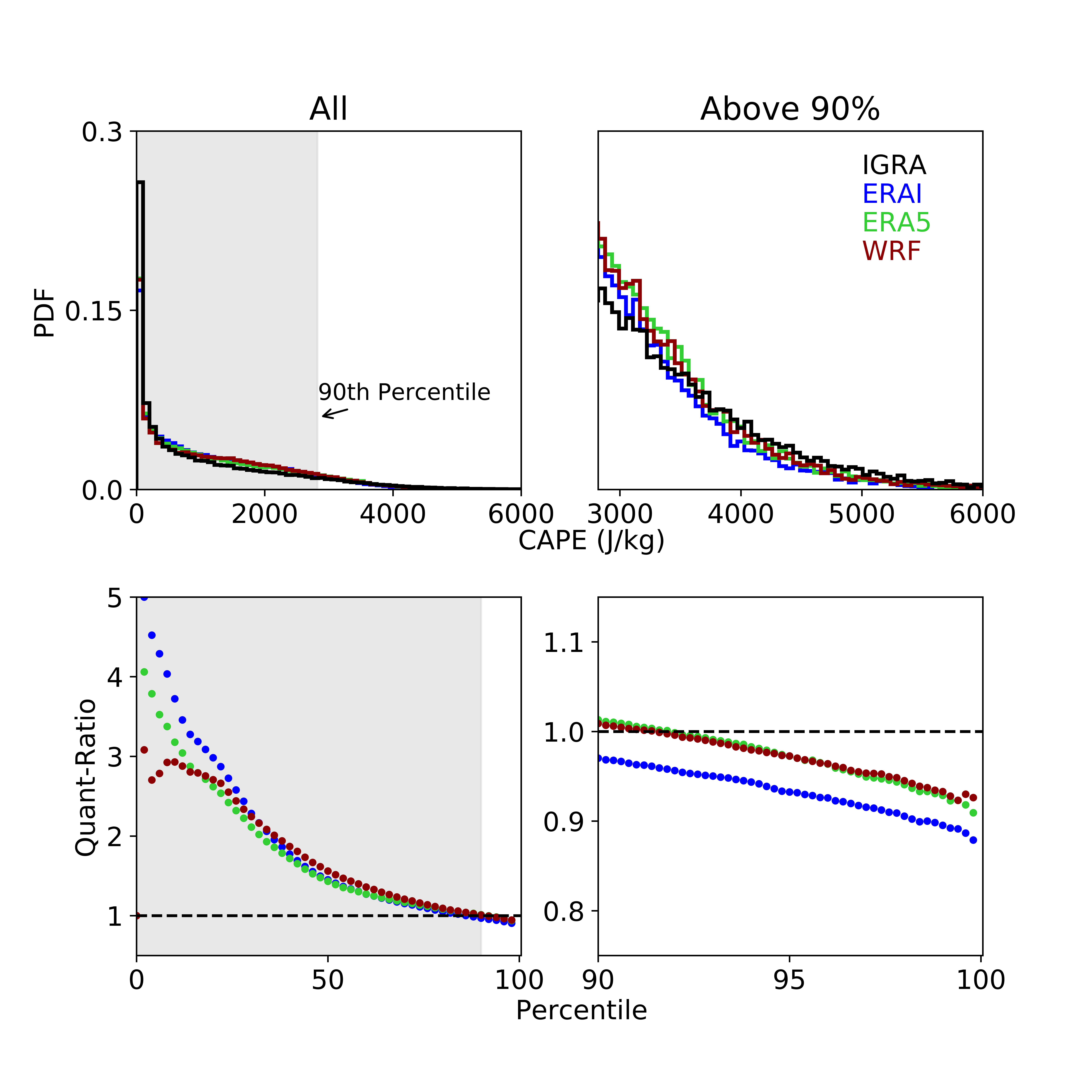}
 \caption{Probability density functions (top row) and quantile ratio plots (bottom row) of CAPE from reanalysis (ERAI and ERA5), high-resolution model output (WRF), and radiosonde observations (IGRA) for MJJA 2001-2012, with times and locations matched to IGRA observations. Points with zero CAPE are excluded (36-40\% of datasets, see Table \ref{tab:distributionSB}). Left column shows full distribution and right column the high tail (90th percentile and above). For IGRA, the 90th percentile is $\sim$2800 J/kg, the 95th $\sim$3200 J/kg, the 97.5th $\sim$4000 J/kg.  In PDFs (top), plots are cut off at 6000 J/kg on the x-axis, omitting less than 0.1\% of all points. The most extreme values are: IGRA (6834 J/kg), ERAI (5198), ERA5 (6291), WRF (7294).  In quantile ratio plots (bottom), a slope downward to the right indicates a narrower distribution.  Model and reanalyses consistently underpredict CAPE values in this high tail. 
}
 \label{fig:SBCAPEdist}
\end{figure}

\begin{figure}[t]
 \includegraphics[width=\linewidth,trim={0.5cm 1cm 1cm 1cm},clip]{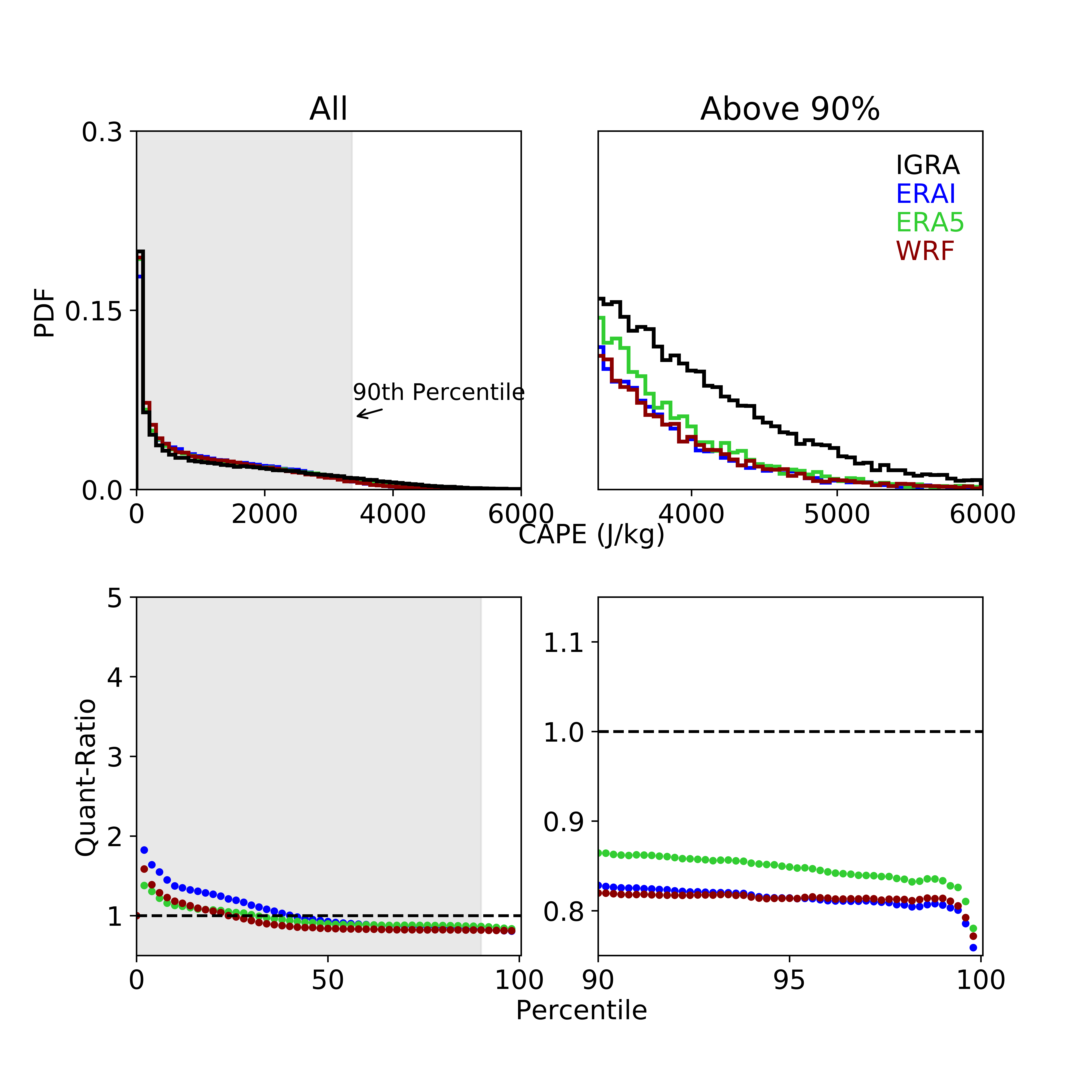}
 \caption{As Figure \ref{fig:SBCAPEdist}, but for MUCAPE instead of SBCAPE. Points with  zero  CAPE  are  excluded  from  the  analysis  (23-35\%  of  the  datasets,  see Table \ref{tab:distributionMU}).  We match the time and locations of model output to IGRA observations.  PDF x-axes are cut off at 6000 J/kg, as less than 0.4\% of all points lie above the limit. For IGRA, the 90th percentile is about 3360 J/kg, the 95th percentile $\sim$4000 J/kg, and the 97.5th percentile $\sim$4540 J/kg.
 \label{fig:MUCAPEdist}
}
\end{figure}

These biases in the high tail are related to a too-narrow distribution of CAPE in model and reanalyses. That is, reanalyses and model produce too few incidences of both extremely low and extremely high CAPE and too many incidences of intermediate CAPE. Figures \ref{fig:SBCAPEdist} and \ref{fig:MUCAPEdist} show distributions of non-zero CAPE values for SBCAPE and MUCAPE, respectively. Because valid zero values make up a large fraction of soundings, the choice whether to include them can potentially affect analysis, but in the datasets here, zero incidences are similar (Tables \ref{tab:distributionSB}--\ref{tab:distributionMU}).  We use two methods to show distributions: histograms (probability density functions, or PDFs) and quantile ratio plots. PDFs provide a basic sense of the CAPE distribution, and quantile ratio plots compare individual quantiles of two distributions to highlight distributional differences. In a quantile ratio plot, simple multiplicative transformation produces a horizontal line whose value is the ratio of means, and a narrowing produces a slope downward to the right.

\begin{table}[h!]
\begin{center}
\resizebox{\linewidth}{!}{\begin{tabular}{| c | c | c | c | c |} 
\hline
& \multicolumn{1}{c|}{IGRA} & \multicolumn{1}{c|}{ERAI} & \multicolumn{1}{c|}{ERA5} & \multicolumn{1}{c|}{WRF}\\ 
\hline
Zeroes & \multicolumn{1}{c|}{36.2\%} & \multicolumn{1}{c|}{38.1\%} & \multicolumn{1}{c|}{35.9\% } & \multicolumn{1}{c|}{39.8\%} \\ 
\hline
$>$ 2000 J/kg & 11.4\% \ & 11.9\% \ (1.04) & 13.3\% \ (1.17) & 12.2\% \  (1.07) \\
\hline
$>$ 3000 J/kg & 4.1\% \ & 3.9\% \  (0.96) & 4.7\% \  (1.14) & 4.2\% \ (1.02) \\
\hline
$>$ 4000 J/kg & 1.6\% \ & 1.3\% \ (0.76) & 1.2\% \ (0.74) & 1.1\% \  (0.68) \\
\hline
\end{tabular}}  \\
\caption{Fraction of observations of SBCAPE in each dataset that exceed threshold values, or have zero value. Data used is the full 2001-2012 MJJA dataset, inclusive of zeroes, with time/location matched to radiosonde observations. Parentheses show the ratio of incidences observed for each model or reanalysis relative to IGRA radiosondes; a number smaller than 1 means underestimation. Note the large deficits in the most extreme SBCAPE category ($>$4000 J/kg), with the number of incidences underestimated by $\sim$25--30\%.
}
\label{tab:distributionSB}
\end{center}
\end{table}

\begin{table}[h!]
\begin{center}
\resizebox{\linewidth}{!}{\begin{tabular}{|c | c | c | c | c |} 
\hline
& \multicolumn{1}{c|}{IGRA} & \multicolumn{1}{c|}{ERAI} & \multicolumn{1}{c|}{ERA5} & \multicolumn{1}{c|}{WRF}\\ 
\hline
Zeroes & \multicolumn{1}{c|}{22.8\%} & \multicolumn{1}{c|}{32.5\%} & \multicolumn{1}{c|}{30.9\% } & \multicolumn{1}{c|}{35.0\%} \\ 
\hline
$>$ 2000 J/kg & 22.2\% \ & 15.9\% \ (0.71) & 16.9\% \ (0.76) & 14.6\% \  (0.66) \\
\hline
$>$ 3000 J/kg & 10.8\% \ & 5.7\% \  (0.53) & 6.4\% \  (0.59) & 5.2\% \ (0.48) \\
\hline
$>$ 4000 J/kg & 3.9\% \ & 1.3\% \ (0.33) & 1.3\% \ (0.35) & 1.1\% \  (0.29) \\
\hline
\end{tabular}} \\
\caption{As in Table \ref{tab:distributionSB} but here for MUCAPE. Deficits in the high tail are larger for MUCAPE than SBCAPE, as expected based on Figure \ref{fig:levelratio}. Parentheses show the ratio of incidences observed for each model or reanalysis relative to IGRA radiosondes. The number of incidences of MUCAPE above the conventional severe-weather threshold (2000 J/kg) is underestimated by $\sim$25--35\% and that of extreme MUCAPE  ($>$ 4000 J/kg) by $\sim$65--70\%. }
\label{tab:distributionMU}
\end{center}
\end{table}

 Reanalyses and model output considered here show the downward and rightward slope characteristic of too-narrow distributions: values are too large in low quantiles and too small in high quantiles.  Above the 95th percentile, SBCAPE quantiles are underestimated by around 5--10\%.
These distributional errors occur even though mean SBCAPE values are similar in all datasets: within +2 to +6\% with zeroes included. In MUCAPE, reanalyses and model have not only narrower distributions but also significant low mean bias,  -22 to -28\%, as expected based on Figure  \ref{fig:levelratio}. This low bias leads to stronger deficits in the high tail, with MUCAPE quantiles above the 95th underestimated by $\sim$17--20\%. MLCAPE biases are intermediate; see Supplementary Figure S1.  Note that mean values of SBCAPE are very similar in all datasets (in fact slightly \textit{larger} in reanalyses and model than in radiosondes); even severe distributional biases may not be reflected in mean values (Supplementary Table S1).

\subsection{Spatiotemporal structure} 
\begin{figure}[h!!]
 \includegraphics[width=0.95\linewidth,trim={0 0.38cm 0 2.2cm},clip]{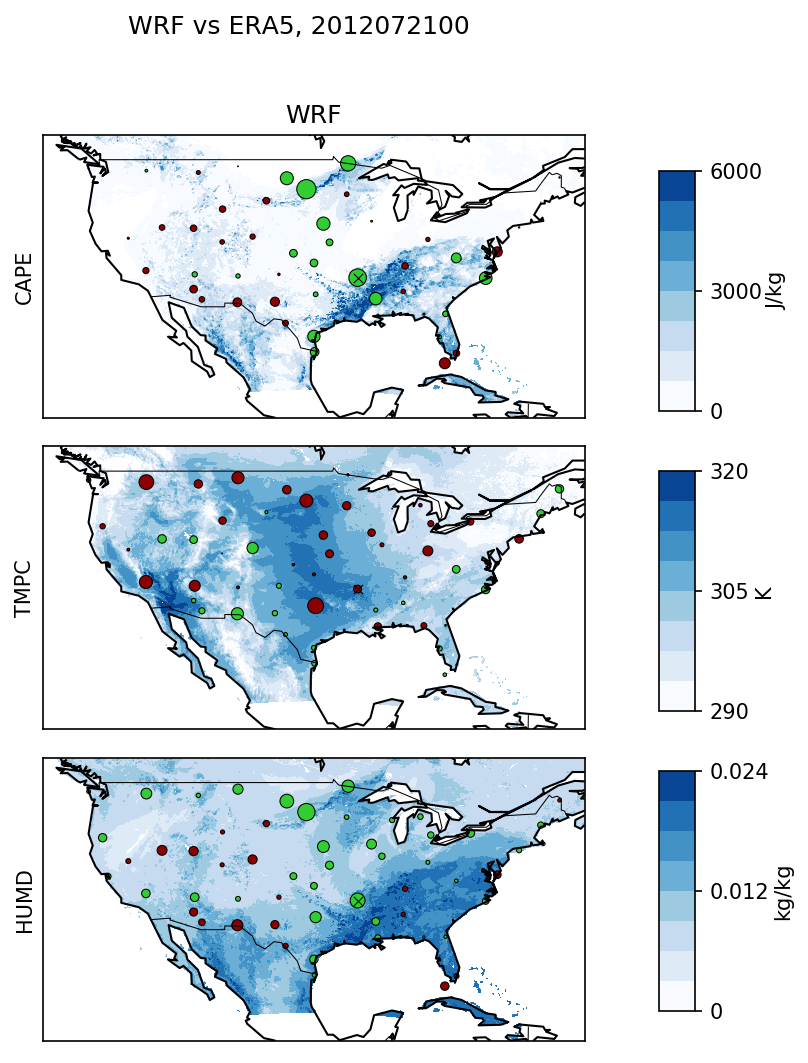}
 \caption{Snapshot of WRF simulation output at 00 UTC, July 21st, 2012. Panel colors show SBCAPE, 2 m temperature, and specific humidity. Ocean values are masked out.  Circles show IGRA stations, with circle area showing the magnitude of bias in each variable and color indicating its sign  (red = high, green = low). Note the low CAPE bias in the Central U.S.\ associated with too hot and too dry model conditions. Sounding marked ``X'' may also be affected by errors in the location of the warm front.}
 \label{fig:spatial}
\end{figure}

Biases might be expected to show spatiotemporal structure, since CAPE is strongly linked to spatially complex fields of temperature and humidity. This relationship is illustrated in Figure \ref{fig:spatial}, which shows a summertime snapshot of surface values from the WRF simulation (SBCAPE, temperature, and specific humidity), coincident with the radiosonde launch time at which CAPE values are typically highest (00 UTC, late afternoon or early evening in the contiguous U.S.). 
The time period shown is affected by a frontal system that brings high humidity to the Southeast and high temperatures to the Central U.S. (See Supplementary Figure S3 for a weather map.) CAPE reaches extreme values only where both temperature and specific humidity are high, resulting in strong spatial gradients and a narrow band of extreme CAPE extending from SE Texas to N.\ Mississippi.

CAPE biases show different spatiotemporal patterns. In Figure \ref{fig:spatial}, two  processes appear to drive spatially correlated CAPE biases:  large-scale patterns of model bias, and mismatches in the location of fronts or other weather features associated with strong gradients.  The former is most striking in Figure \ref{fig:spatial}: the model is too warm and too dry in the Central U.S., coincident with and likely causing a large region of underestimated model CAPE. The warm-and-dry bias in this WRF simulation is extensively documented \citep{liu_continental-scale_2017, morcrette_introduction_2018}.  

Large-scale and weather-related biases have different consequences for CAPE comparisons with observations. Large-scale biases should be persistent, and will affect the overall distribution of CAPE. Fine-scale weather-related errors vary rapidly on timescales of hours and should have minimal distributional effect, but will produce severe mismatch in individual soundings. That is, even if the overall distribution is well-captured, models and reanalyses can fail to accurately represent observed CAPE at every time step. Inaccuracy in location or timing of weather phenomena can produce considerable scatter in a comparison of individual soundings.

\subsection{Calibration with ground observations}

Scatter in SBCAPE errors is in fact large in the model and reanalysis products considered here, with correlation coefficients against radiosonde values of only R = 0.74--0.86. Figure \ref{fig:scatter} shows the comparison of WRF and radiosondes (top left, R = 0.74); see Supplementary Figures S4--S5 for ERAI and ERA5.
Similar behavior is found in other studies, e.g.\ \citet{gensini_severe-thunderstorm_2013} who found correlation coefficients of 0.36--0.71, and  \citet{gartzke_comparison_2017}, who show that reanalysis and satellite pseudo-soundings cannot reproduce radiosonde observed SBCAPE at individual timesteps.

\begin{figure}[h!]
 \includegraphics[width=\linewidth]{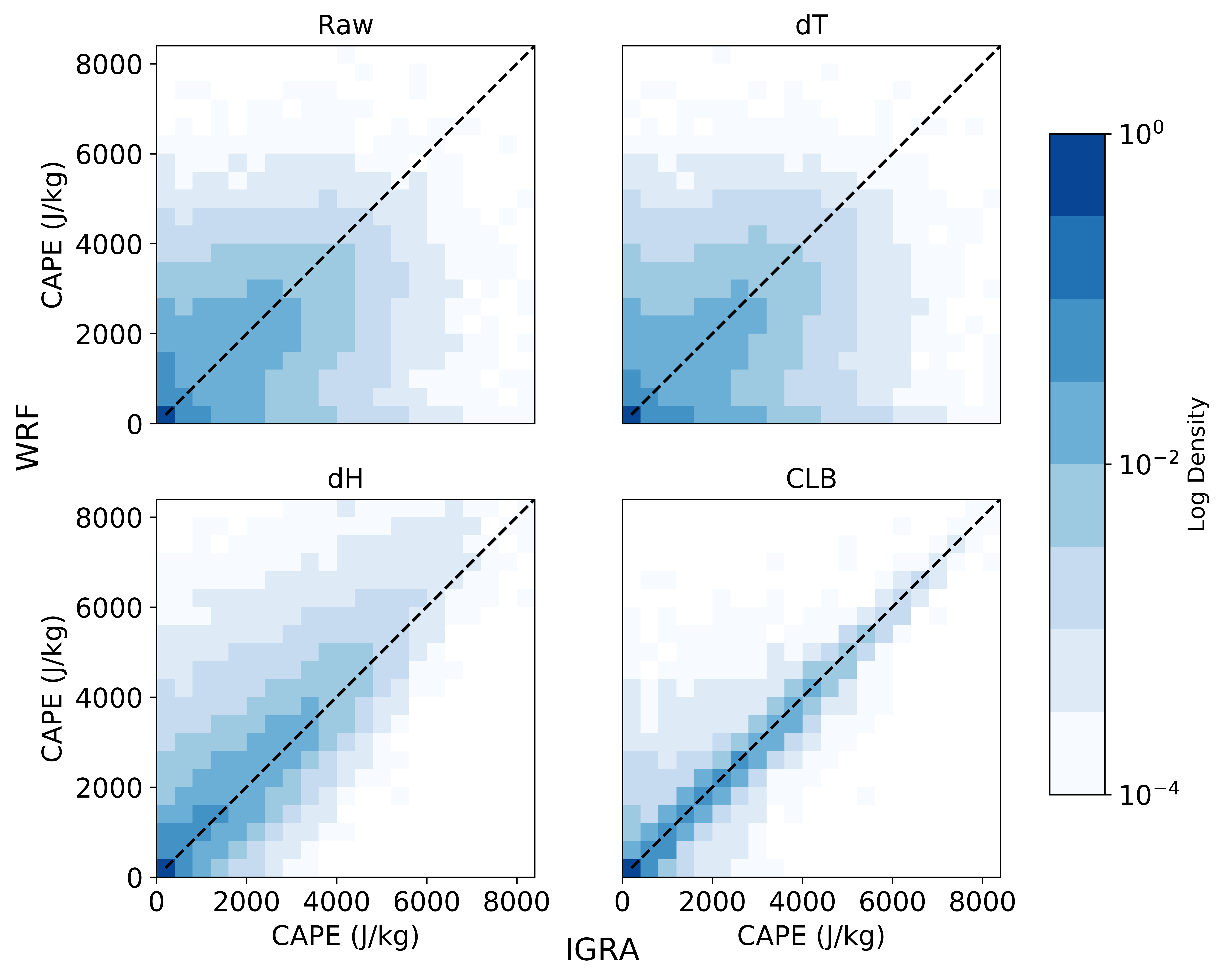}
 \caption{Comparison of SBCAPE in WRF and radiosonde observations, for all points during summer (MJJA) 2001-2012 when observations are available, inclusive of zeroes.  Color bar shows log density (midpoint color is 1\% of all observations), and both axes are also log scale.  ({\bf Top left}) Raw data, showing wide scatter. Other panels: Recalculated WRF CAPE using ({\bf top right}) observed surface temperature,  ({\bf bottom left}) observed surface humidity, and ({\bf bottom right}) all surface values from observations. All recalculated CAPE values also involve a pressure correction whose effects are small. For analogous figures for ERAI and ERA5, see Supplementary Figures S4--S5.}
 \label{fig:scatter}
\end{figure}

Following \citet{gartzke_comparison_2017}, we test to see if these inaccuracies can be corrected by simply replacing surface thermodynamics fields with those from radiosondes (Figure \ref{fig:scatter}). That is, we test whether errors in model and reanalysis SBCAPE are driven primarily by surface conditions rather than by the structure of atmospheric profiles. Both factors can be important because CAPE is a function of the integrated buoyancy across the convective layer, which is determined by both parcel and environmental temperature and moisture. In Figure \ref{fig:scatter}, we successively replace surface values in WRF output, first temperature and pressure (top right), then specific humidity and pressure (bottom left), then all surface fields (bottom right). 

Surface values do seem to govern SBCAPE bias almost entirely. For WRF, correcting the surface specific humidity raises the correlation coefficient from 0.74 to 0.92, and replacing all surface fields raises it to 0.99, removing scatter almost entirely.  While correcting temperature does not raise the correlation coefficient in WRF, and instead lowers it to 0.71, for other datasets the temperature correction also contributes positively; see Supplementary Table S2. We also consider an alternate measure of correspondence, the percentage of points that fall within  $\pm 800$ J/kg of the one-to-one line (the width of two cells in Figure \ref{fig:scatter}). For raw WRF data, the percentage is 72.7\% (RMSE = 847 J/kg); correcting surface temperature raises the percentage slightly to 73.7\% (RMSE = 875 J/kg); correcting surface humidity  raises it to 84.5\% (RMSE = 539 J/kg), and full calibration to 99\% (RMSE = 169 J/kg). Results for ERAI and ERA5 are similar. Adjustment of surface values also largely corrects the distributional problems at high CAPE, so that for quantiles above 0.9, corrected SBCAPE values in reanalyses and model match those from radiosondes to within -0.4\% to +1.8\%. (Compare Figure \ref{fig:SBCAPEdist} with Supplementary Figure S2).

\section{Results -- CAPE in temperature \& humidity space}
The fact that reanalyses and modeled SBCAPE can be brought into agreement with radiosondes by simply replacing surface values implies that thermodynamic fields at upper levels are not important factors in SBCAPE biases. It may then be reasonable to consider SBCAPE as a function of surface  thermodynamic fields alone. We therefore examine SBCAPE in the 2D parameter space of temperature (T) and specific humidity (H) to ask:
1) Is the density distribution of SBCAPE in T--H parameter space similar in reanalyses, model, and radiosondes? 2) What surface conditions are related to the highest SBCAPE days? and 3) What factors drive model and reanalysis biases in SBCAPE?

\subsection{Dependence on surface temperature and humidity}
\begin{figure}[t]
 \includegraphics[width=\linewidth]{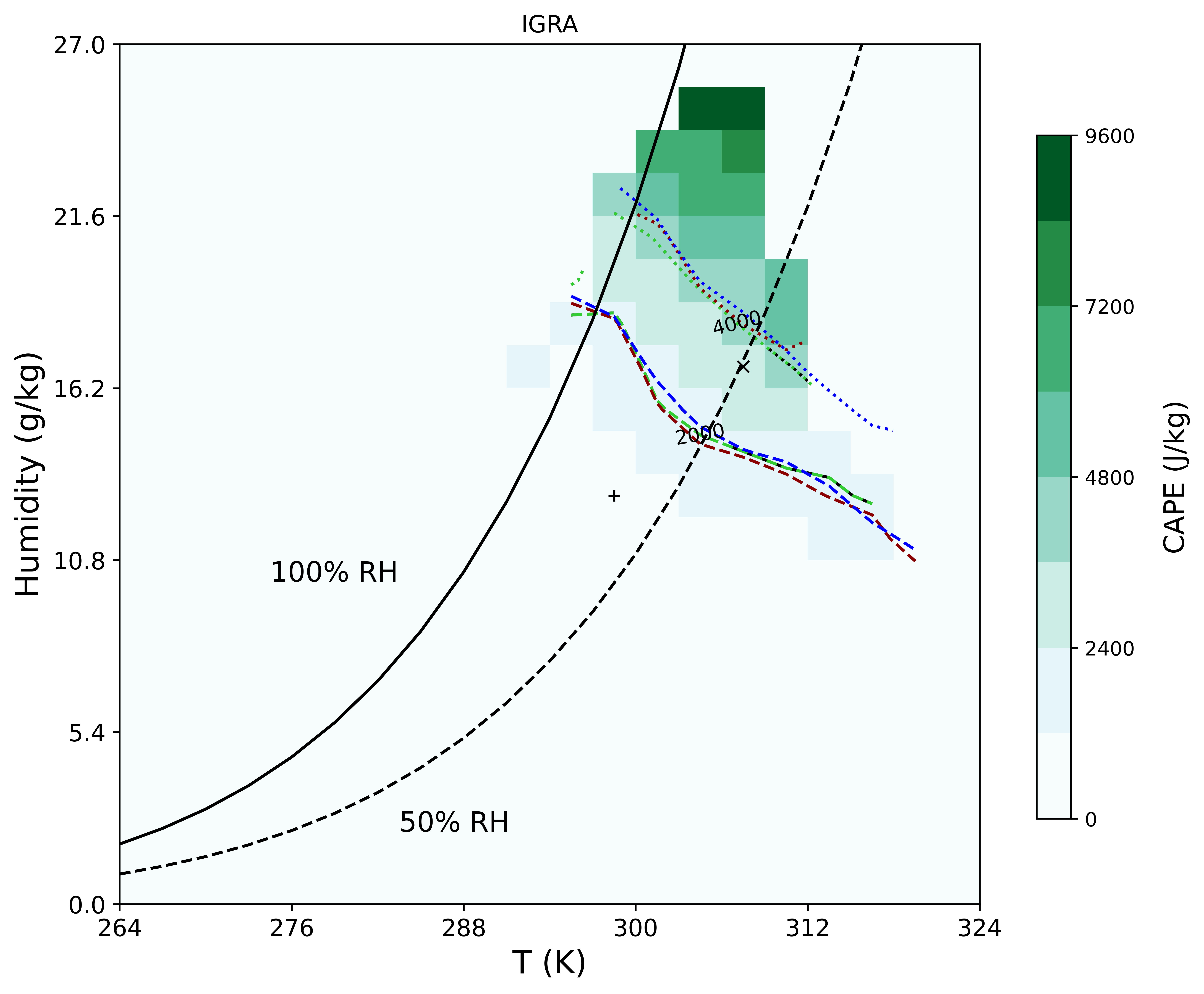}
 \caption{Mean radiosonde observed SBCAPE in surface temperature and specific humidity parameter space, for the entire dataset: summer (MJJA) 2001--2012 over the contiguous U.S., inclusive of all launch times and of zero values. Colors denote mean CAPE values averaged in bins of 3 K and 1.35 g/kg. Solid and dashed lines mark conditions of 100\% and 50\% relative humidity. Symbols '+' and 'x' mark two cases ('warm' and 'hot') used in Figure \ref{fig:bins}. Contours show approximate limits for 2000 and 4000 J/kg SBCAPE for all datasets with no surface corrections applied: IGRA (black), ERAI (blue), ERA5 (green), and WRF (red). Similarity of contours means that all datasets show similar bivariate distributions. See Supplementary Figure S6 for full density distributions for all datasets.}
 \label{fig:heatmapcape}
\end{figure}

CAPE distributions in T--H parameter space are in fact highly robust across all datasets. Figure \ref{fig:heatmapcape} shows the heatmap of mean CAPE for radiosonde measurements, with data binned in steps of 3 K and 1.35 g/kg. CAPE values show a smooth gradient from lowest values at bottom left (warm and dry conditions) to highest at top right (hot and humid). Contour lines at 2000 and 4000 J/kg for radiosonde observations are nearly identical to those for other datasets (overlain). This similarity means that surface T and H robustly predict SBCAPE in all datasets, and supports the previous finding that bias in SBCAPE can be explained by bias in surface measurements alone.

Only a restricted set of conditions tend to produce the high CAPE values associated with extreme weather. The 2000 J/kg contour is a commonly used threshold for severe weather, first defined by \citet{brooks_spatial_2003} and subsequently used in multiple studies of future changes (e.g.\ \citealt{trapp_transient_2009, diffenbaugh_robust_2013}).  In all datasets the conditions producing mean SBCAPE above this threshold involve temperatures above 297 K for 100\% relative humidity (RH), or above 304 K for 50\% RH. For the 4000 J/kg threshold we use in this work, the required temperatures are 2--3 K warmer, 299K at 100\% RH or 307K at 50\% RH. Significantly higher SBCAPE values are possible: in the most extreme conditions regularly sampled by radiosonde, 308 K at 65\% RH, the average observed SBCAPE is over 7400 J/kg.  Reanalyses and model rarely produce SBCAPE values this high (0.0008\% of incidences, while observed incidences are nearly 10x more frequent at 0.006\%) not because they differ in fundamental atmospheric physics but because they rarely sample the appropriate surface conditions. (See also Supplementary Figure S6.)

\begin{figure}[t]
 \includegraphics[width=\linewidth,trim={0 1cm 0 0.5cm},clip]{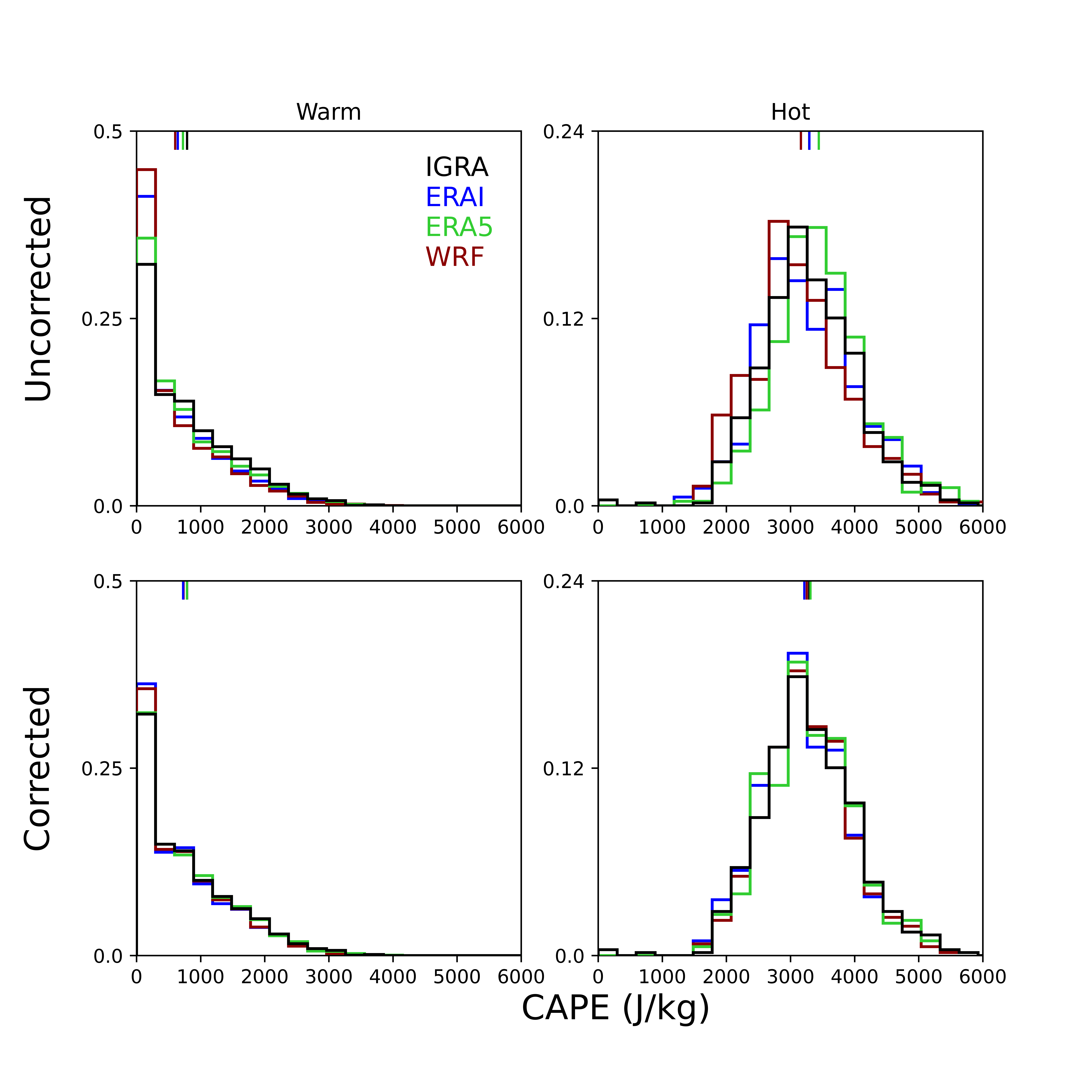}
 \caption{Comparison of SBCAPE in all datasets for specified T--H grid cell: the `Warm' example is centered at 298.5 K and 12.825 g/kg (63.4\% RH), and has mean SBCAPE 810 J/kg; the `Hot' example is at 307.5 K and 16.875 g/kg (48.7\% RH) with mean SBCAPE 3270 J/kg. Each bin is 3 K in width, and 1.35 g/kg in height. Top row shows uncorrected SBCAPE from reanalyses and model, and bottom corrected with IGRA surface values. Note that since the correction involves adjusting surface T and H, the profiles sampled in top and bottom rows are different. The `warm' bin has 390 profiles in the uncorrected data and 480 in the corrected, while `hot' has 2300 and 2111, respectively.  Tickmarks at panel top show the mean of each distribution. Distributions are very similar; correcting surface values only slightly adjusts means (from a maximum bias of -5\% in uncorrected data to -2\% after correction).}
 \label{fig:bins}
\end{figure}

While the heatmap of Figure \ref{fig:heatmapcape} describes average CAPE across all incidences, each T--H grid cell involves an underlying distribution. It is valuable to ask whether these underlying distributions are  also similar across datasets. Figure \ref{fig:bins} highlights distributions in two example grid cells differing in temperature by 9K, one representing hot conditions with high mean surface CAPE (3270 J/kg) and the other cooler conditions of lower mean CAPE (810 J/kg).
The resulting distributions are radically different: the `warm' cell has a left-skewed distribution with a mode of zero, while the `hot' cell distribution is near-normal. For each set of surface conditions, however, distributions are highly similar across datasets, not only in the corrected profiles (bottom row) but even in the uncorrected ones (top row), which sample different individual soundings. That is, CAPE distributions based on profiles of similar surface T--H are similar even when profiles have the wrong surface values and are assigned to the wrong T--H grid cell. The correction produces a small shift in the mean, but even biased surface values appear highly predictive of the distribution of calculated SBCAPE. (For distributions of T and H error in these cases, see Supplemental Figure S7 and Table S3.)

\subsection{Identifying sources of CAPE bias} 
\begin{figure}[t]
 \includegraphics[width=\linewidth]{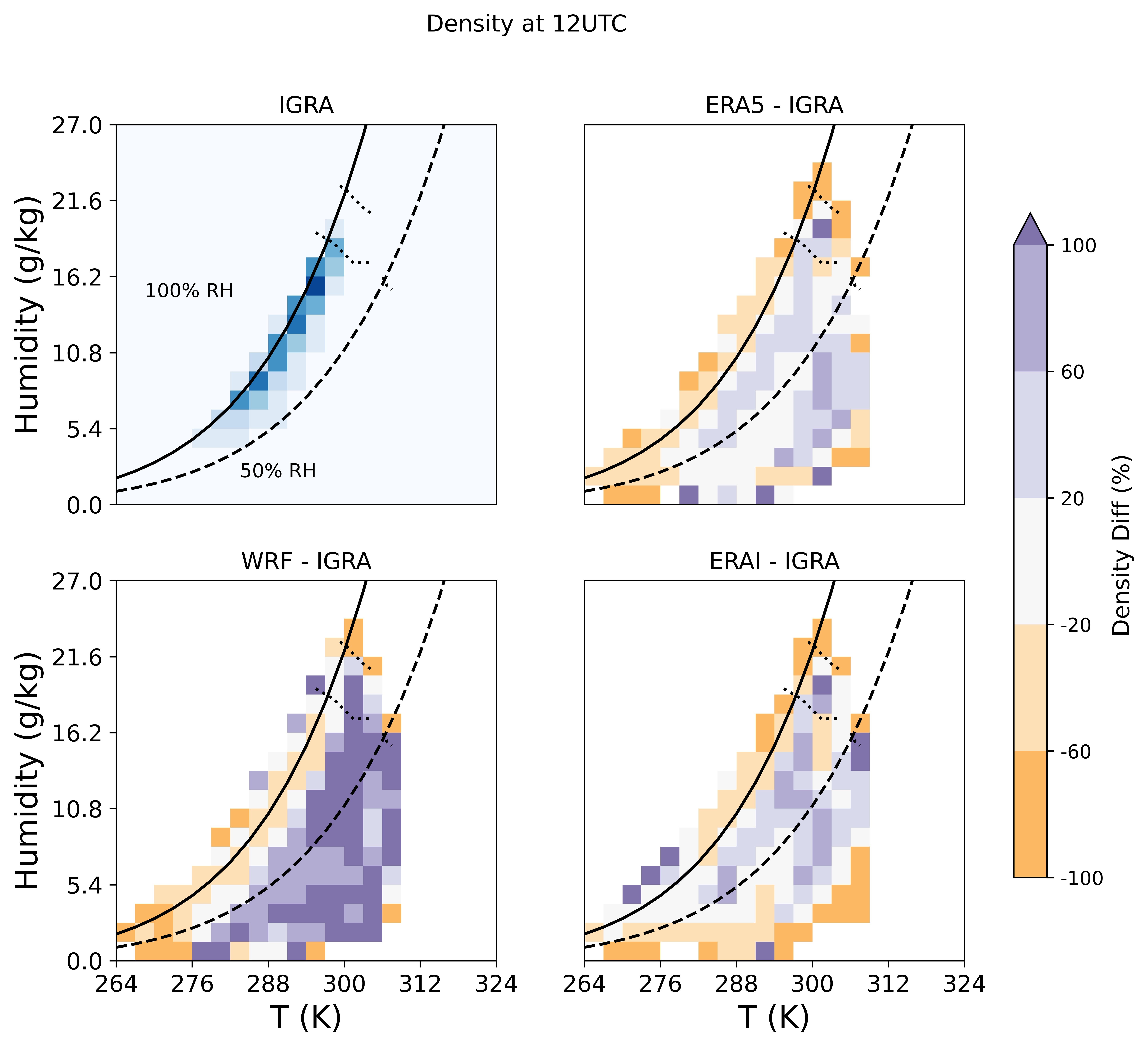}
 \caption{\textbf{(Top left)} Density of observed surface conditions in temperature -- specific humidity parameter space at 12 UTC (early morning in the contiguous U.S.), again for 2001-2012 MJJA radiosonde observations. Contours are repeated from Figure \ref{fig:heatmapcape} to mark conditions associated with 2000 and 4000 J/kg SBCAPE.  
 Darkest blue color shown is 5.6\%--6.4\% of distribution; lightest is 0--0.8\%. Grids with no more than three samples are defined as outliers and removed (only 0.03\% of all model or reanalysis samples). Nighttime and early morning conditions are tightly distributed in relative humidity (RH $\sim$80\%) and tend to be relatively cool (T $<$ 300 K), with almost no conditions sampled that would tend to produce SBCAPE $>$2000 J/kg. \textbf{(Other panels)} --  heatmaps of density differences between models and observations for (clockwise from top left) ERA5, ERAI, and WRF. Color scale shows fractional difference after normalizing each bin by IGRA raw density. Orange = undersampling and purple = oversampling.  Reanalysis and model all underestimate relative humidities (orange near the RH=100\% contour) and WRF shows a strong warm dry bias (dark purple in lower right). }
 \label{fig:heatmaputc12}
\end{figure}

\begin{figure}[t]
 \includegraphics[width=\linewidth]{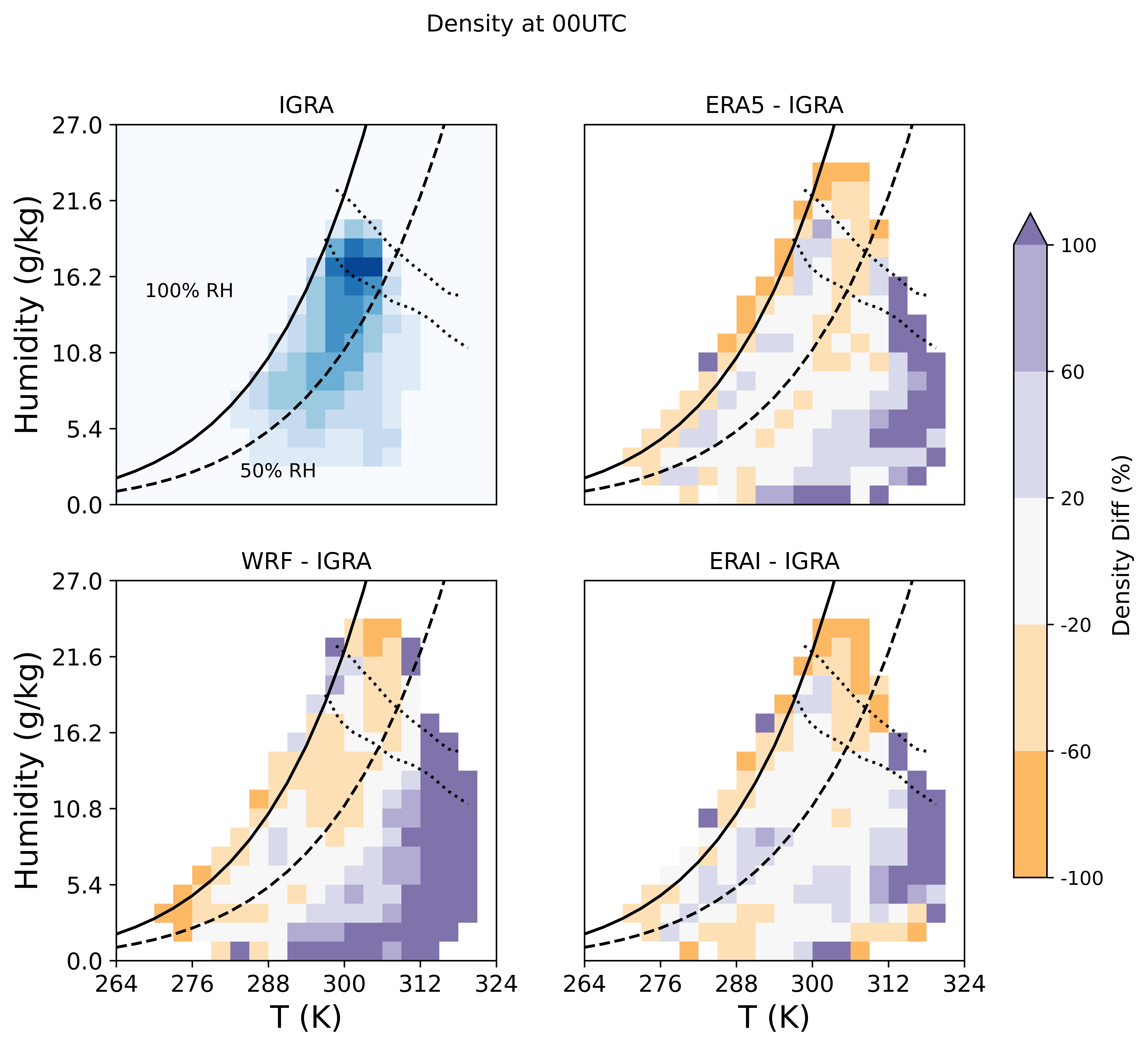}
 \caption{As in Figure \ref{fig:heatmaputc12}, but for 00 UTC (late afternoon / early evening in the contiguous U.S.) \textbf{(Top left)} Density of observed surface conditions in T--H diagram. At this time period the density distribution peaks in conditions associated with 2000--4000 J/kg CAPE. Darkest blue color shown is 2.1\%--2.4\% of distribution; lightest is 0--0.3\%.  \textbf{(Other panels)} Density differences between renalyses / model and radiosondes. ERA5 and ERAI undersample both the highest relative humidities and the highest temperatures (orange near the RH=100\% contour and on the right side), while WRF shows a warm dry bias (purple in lower right). Reanalyses and model all severely undersample the conditions associated with extreme CAPE (orange in upper right).} 
 \label{fig:heatmaputc00}
\end{figure}

Because surface temperature and humidity are predictive of SBCABE, biases in SBCAPE in reanalyses and model appear driven by biases in these surface thermodynamic values. We can therefore use the T--H diagram to identify the factors that lead to underprediction of the high tail of CAPE. Figures \ref{fig:heatmaputc12} and \ref{fig:heatmaputc00} use the same T--H diagram as in Figure \ref{fig:heatmapcape}, only now we show not the heatmap of CAPE but the density of observations of each T--H grid cell and the difference in that number between datasets. Because the diurnal cycle strongly affects surface values, we show separate figures for 00 UTC (U.S.\ late afternoon/evening) and 12 UTC (U.S.\ early morning). At both time periods, biases in ERAI and ERA5 are similar in character, and those in WRF are distinct: that is, their joint distributions of surface values omit different parts of the T--H space. Reanalyses and model all underestimate the extreme T--H values associated with extreme CAPE, but they may do so for different physical reasons.

Of the two times routinely sampled by radiosondes, the cooler 12 UTC launches do not generally involve conditions associated with high CAPE (Figure \ref{fig:heatmaputc12}, upper left).  These early morning conditions show the influence of nighttime cooling, and almost no conditions experienced would tend to produce SBCAPE $>$2000 J/kg.  (Nearly all observations in the `warm' example of Figure \ref{fig:bins}  come from 12 UTC.)  In this time period, observations are all fairly high in relative humidity, with a tight distribution centered around $\sim$80\%. Reanalyses and model have both low bias in highest RH and, for ERA, a still narrower distribution: all datasets underpredict incidences of RH close to or above saturation and ERA also underpredicts those significantly below, while WRF overpredicts warm dry soundings (Figure \ref{fig:heatmaputc12}, remaining panels). This combination of warm and dry bias explains why correcting WRF surface temperatures alone does not improve the match to radiosonde CAPE measurements. The dry bias (low RH) issues appear driven by temperature biases; see Supplementary Figures S8--S10, bottom panels, for absolute biases in T--H space at 12 UTC.

The 00 UTC launches show the result of daytime warming (Figure \ref{fig:heatmaputc00}, upper left), with temperatures warmer than at 12 UTC. Because specific humidity does not change much, relative humidities are considerably lower in the late afternoon 00 UTC soundings. This time period includes most of the high CAPE values sampled, and the modal (most probable) surface conditions sampled have mean SBCAPE $\sim$3000 J/kg (between 303--309 K and $\sim$50\% RH, similar to the `hot' example of Figure \ref{fig:bins}). As at 12 UTC, ERA reanalyses undersample the highest and lowest relative humidities and omit the highest temperatures almost completely (Figure \ref{fig:heatmaputc12}, remaining panels). WRF again undersamples cold and humid conditions but oversamples hot and dry ones. Both model and reanalyses therefore fail to capture the extreme hot and humid conditions associated with the highest CAPE levels. See Supplementary Figures S8--S10, top panels, for absolute biases at 00 UTC.

%\newpage
\section{Results -- diurnal cycles of CAPE and biases} 
The dependence of SBCAPE on surface temperature and specific humidity means that biases in CAPE are largely determined by biases in the diurnal cycle of surface thermodynamic fields. We therefore examine the timing and amplitude of the diurnal cycle in reanalyses, model, and radiosondes to determine whether any consistent features underlie the biases described above.

\begin{figure}[t]
 \includegraphics[width=\linewidth]{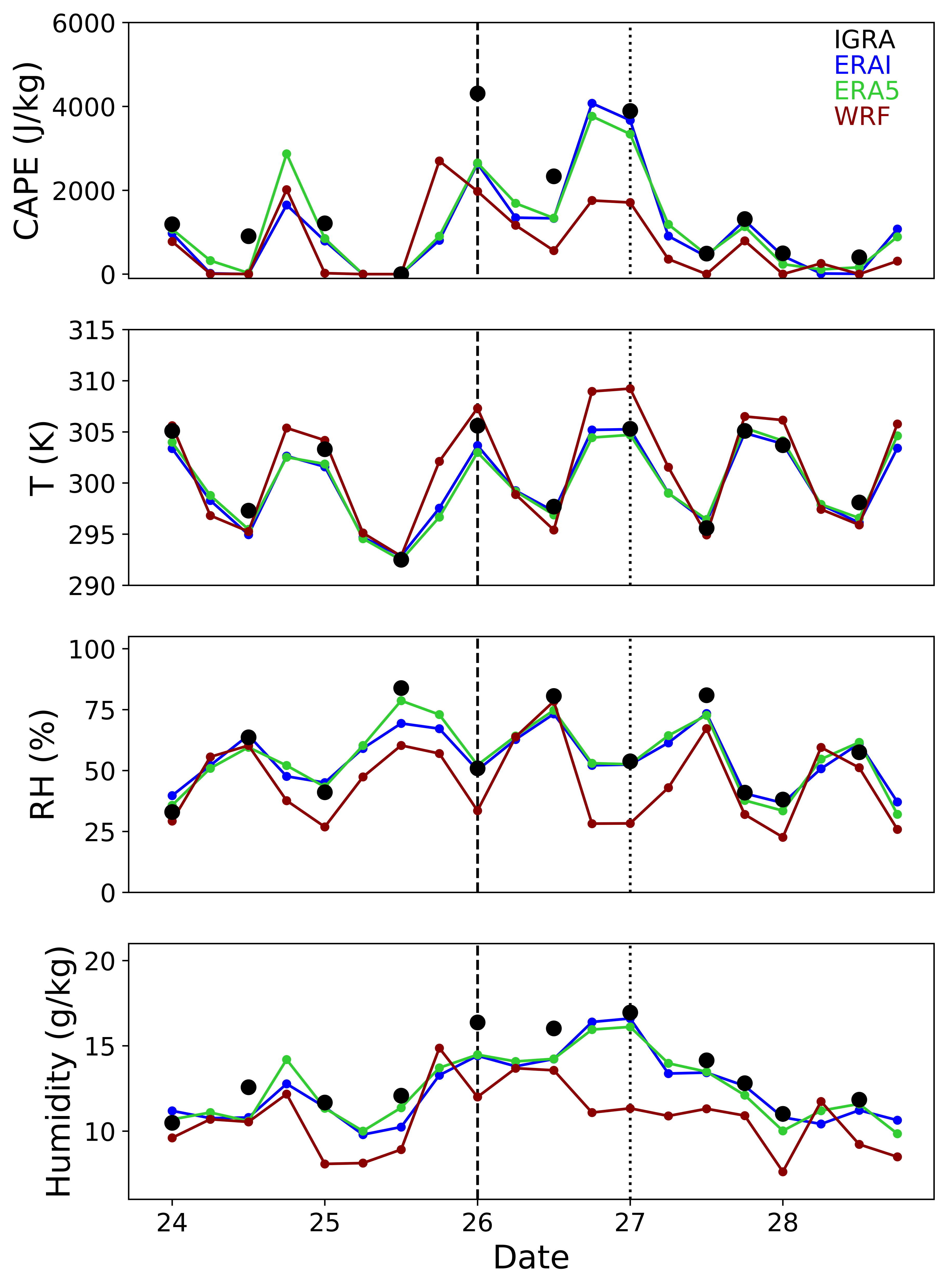}
 \caption{An example episode of high CAPE and substantial CAPE error: 5 days from May 24th to 28th, 2012 over Topeka, Kansas, color coded as usual. Reanalyses and model are shown at 6-hourly; IGRA soundings are every 12 hours (an additional records at 27th 18 UTC). Vertical lines mark the two examples discussed in text. CAPE biases arise from underestimated specfic humidity: in ERA renalyses on May 26 because T is low despite accurate RH, and in WRF on both May 26 and 27 because conditions are too dry even though temperature is high.} 
 \label{fig:episode}
\end{figure}

To illustrate diurnal variations and show explicitly how CAPE evolves, we show in Figure \ref{fig:episode} an episode exhibiting large CAPE error, which is broadly representative of problematic reanalyses and model pseudo soundings. Figure \ref{fig:episode} shows a 5-day sequence from May 24th--28th, 2012 at a station in Topeka, Kansas, which often experiences high summertime CAPE values and strong convection.  All datasets exhibit strong ($\sim$10K) diurnal swings in temperature, but relatively small changes in specific humidity, so that relative humidity falls substantially in daytime as the temperature warms. In radiosondes, mean RH drops from 78\% in the early morning (12 UTC) to 49\% in late afternoon/early evening (00 UTC). (These values are typical for our dataset; summertime mean 00/12 UTC radiosonde RH across the contiguous U.S.\ is 79\%/56\%.)  Around May 26th, the station sees an influx of moister air, and radiosonde profiles show two instances of extreme CAPE, nearly 4000 J/kg at 00 UTC on May 26th and 27th. All reanalyses and model grossly underestimate the May 26th episode by nearly 2000 J/kg; on the following day bias persists in WRF alone. 

\begin{figure}[h!]
\begin{center}
 \includegraphics[width=\linewidth,trim={0.5cm 0.3cm 0.5cm 1cm},clip]{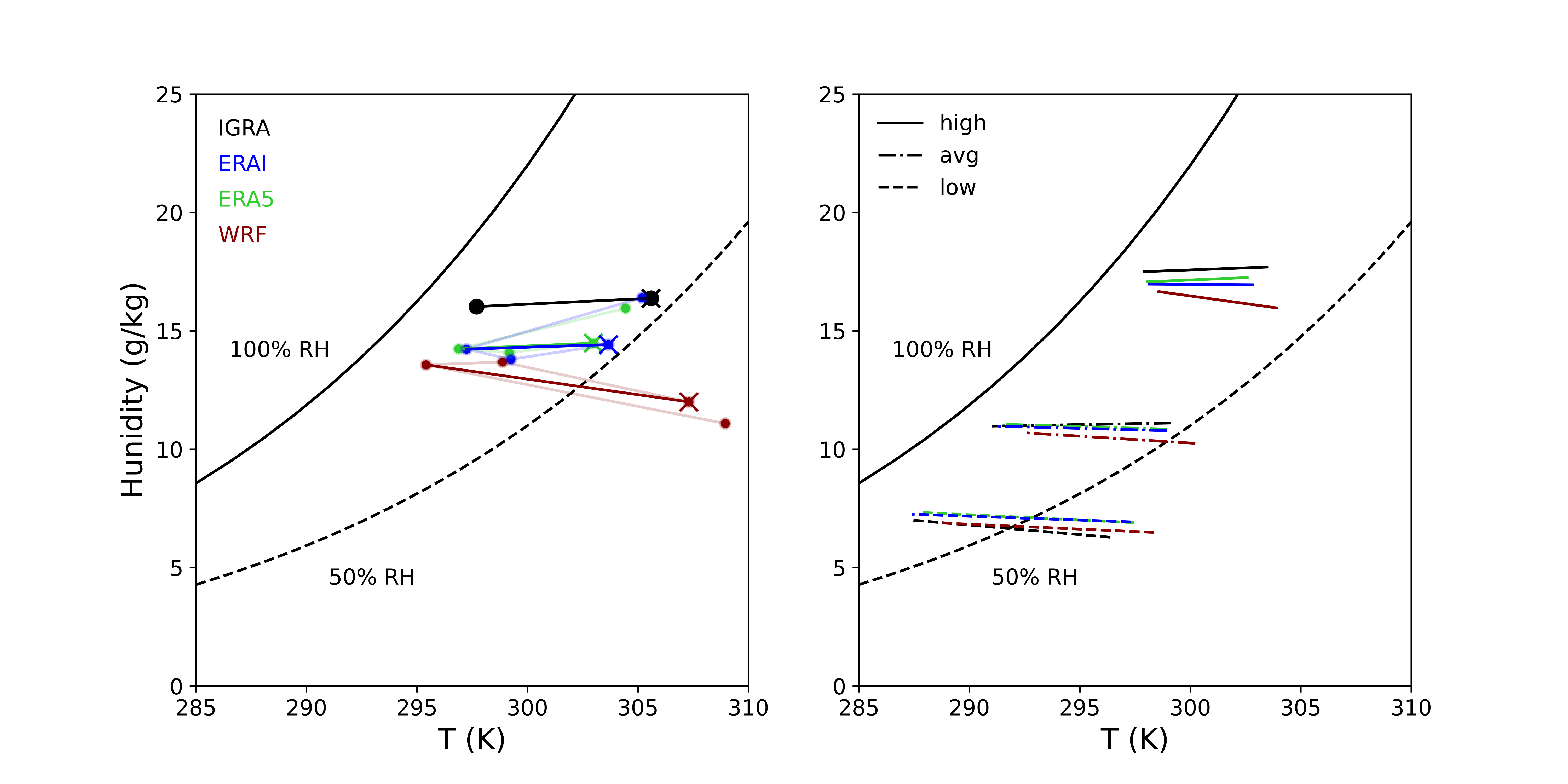}
 \caption{The diurnal cycle in T--H space in all datasets. Color code follows the convention throughout this work. \textbf{(Left)} The example of Figure \ref{fig:episode}. Thick lines connect 00 UTC (right end) and 12 UTC (left end) values on the day of high CAPE error (May 26th 2012), with time of maximum error marked with 'x'. Faint lines connect 6-hourly values in reanalyses and model over the full 5-day period. \textbf{(Right)} Mean summertime diurnal cycles.  Lines connect mean of 00 UTC and 12 UTC points for each dataset, for all profiles (dot-dashed), and for high-CAPE (solid) and low-CAPE (dotted) subsets, defined as 00 UTC SBCAPE values above 90th / below 10th percentile in each dataset, and values 12 hours later. In all cases the IGRA diurnal cycle involves flat or slightly increasing H; the WRF cycle decreasing H; and the ERA cycles too low humidity and too small a daytime T rise in high CAPE conditions.}
 \label{fig:diurnal}
\end{center}
\end{figure}

CAPE biases in the example episode of Figure  \ref{fig:episode} are produced by different behavior in  reanalyses and model, each of which produces a deficit in specific humidity. Early morning temperatures match reasonably well in all datasets, but ERA reanalyses show too-small daytime temperature rise on several days. ERA RH is reasonably accurate throughout, so the too-low peak temperatures are associated with a specific humidity deficit. An example is that ERA underestimates CAPE on May 26th due to a combination of low bias in both temperature and specific humidity. WRF on the other hand shows excessively large daytime temperatures but is substantially too dry in both relative and specific humidity. During the two ``missed-high-CAPE'' episodes, WRF RH is $\sim$25 percentage points below that in radiosondes. The WRF daytime dry bias is amplified because specific humidity often drops during the day, something not seen in reanalyses or radiosondes. Dry bias in WRF dominates the low bias in CAPE, whereas the warm bias does not help.

The diurnal cycle is best visualized using the temperature-humidity (T--H) diagram shown previously. Figure \ref{fig:diurnal} shows diurnal cycles both in the example episode (left) and in climatological mean values over 2000--2012 summers (right).  For climatological summer means, we show not only the overall average but also subsets of days involving the highest and lowest radiosonde SBCAPE values (90th/10th percentiles). The comparison shows that many  features of the May 2012 example episode are typical. In the climatological mean, biases are again smaller during the nighttime and larger during the day. The ERA reanalyses show too-weak daytime warming and are slightly too dry, but with daytime specific humidity rising slightly as in radiosondes. In WRF, the amplitude of the diurnal temperature cycle is approximately correct, but temperatures are biased high by $\sim$1.3 K. WRF is also substantially too dry, and the bias is exacerbated when specific humidity erroneously drops in the daytime.

\section{Conclusion and Discussion}

% Conclusion Paragraph #1: 
%\textit{Para 1 - SUMMARY strong bias in high values can be important for modeling and understanding severe weather. }
Despite the importance of CAPE to both model construction and meteorology, few prior studies have evaluated CAPE biases against radiosondes and none have done so on a large enough scale to evaluate climatological distributions. Distributional biases that affect CAPE extremes can be significant because CAPE plays a direct role in model convective parametrizations, informing both convective triggering and total mass flux, and indirectly affecting precipitation diurnal timing and amplitude. More broadly, CAPE is a key meteorological parameter linking the large scale environment to weather-scale events, and potential increases in warmer climate conditions are of strong scientific interest. Misrepresentation in the present lends some concern about interpreting model projections of future changes. 

This study of nearly 200,000 proximity soundings in two reanalyses and a convection-permitting model confirms consistent patterns of distributional bias. CAPE distributions are too narrow in all reanalyses and model studied here, with undersampling of the most extreme values that are associated with severe weather events. Values in the high tail (95th percentile and above) are 5--10\% too low in surface-based SBAPE and even more severely underestimated at the most unstable layer, at 17--20\% too low in MUCAPE. 

Note that the distributional problems that affect the high tails are not manifested in CAPE means. The datasets studied here actually slightly overestimate mean SBCAPE, by 6--17 J/kg or 2--6\%. Validating mean CAPE is therefore insufficient when using models and data products to understand severe weather and strong convection. Studies that attempt to diagnose biases in extreme CAPE by matching soundings to severe weather events are also insufficient since model displacement of weather events means that ``mismatch'' error is large and proximity soundings will not necessarily capture the same meteorological context.

%\textit{Para 3 - CAPE/bias is strongly dependent on surface.} 
In this study, both distributional biases and ``mismatch error'' in CAPE appear driven by conditions at the surface and/or boundary layer. SBCAPE shows a tight and similar dependence on surface temperature and humidity in all datasets; the dependence is so strong that we can reproduce CAPE distributions as a function of T,H even with the incorrect set of vertical profiles. 
In this study, the too-narrow SBCAPE distribution appears the consequence of too-narrow distributions of surface values, especially in RH space: model and reanalyses undersample both very high and low RH values. This bias curtails the high tail of CAPE associated with hot and humid conditions. Because distributional issues show commonalities across very different data products -- reanalyses with parametrized convection and simulations with resolved convection -- they are necessarily unrelated to the treatment of convection. 

The dependence of SBCAPE on surface conditions has been noted by many previous authors (e.g.\ \citealt{maddox_examination_1982, zhang_quasi_2002, donner_convective_2003, guichard_cloud_2004}), but discussion of improving models to improve CAPE representation have tended to focus on the atmospheric profile. This study emphasizes the importance of the land surface and boundary layer models instead. Misrepresentation of either evaporation from the surface or vertical mixing of the boundary layer can critically affect CAPE. In this study, the greater bias in MUCAPE than SBCAPE across all datasets points to boundary layer processes as common problematic elements. In all datasets, in profiles with significant CAPE the most-unstable layer lies well within the boundary layer: $<$ 50 hPa from the surface in 90\% of profiles with CAPE $>3000$ J/kg. Multiple prior studies have found that boundary layer schemes can modify the diurnal cycle of temperature and humidity through their treatment of mixing \citep{kalthoff_2009, coniglio_verification_2013,garcia-diez_numerical_2013, xu_assessment_2019}. In particular, the strength and persistence of inversions affects the vertical diffusion of moisture in the boundary layer \citep{kalthoff_2009}, and erroneous daytime decrease in specific humidity like that seen in WRF can result from excessive mixing between the boundary layer and the free troposphere \citep{shin_effects_2007}.

%\textit{Para - Future work / implications for future rise.} 
% relationship to surface T,H is stronger than Clausius-Clapeyron
It is important to note that the temperature dependence of CAPE in measured or modeled present-day profiles need not match future changes under CO$_2$-induced warming. In the present-day profiles considered here, an increase of 1 Kelvin in surface temperature at constant RH results in an increase in SBCAPE of $\sim$15--30\% throughout most of the moderate- to high-CAPE regime. (That is, the slope of the response surface of Figure \ref{fig:heatmapcape} along a line of constant RH is 15--30\%/K for CAPE $>$ 2000 J/kg; see also Supplementary Table S5 and Figure S11.) Future CAPE rises are assumed to be smaller. Theoretical considerations suggest that SBCAPE in convective regimes should rise following Clausius-Clapeyron, at 6--7\%/K \citep{romps_clausius_2016}, and modeled changes are roughly similar (e.g.\ $\sim$6\%/K in MLCAPE in the midlatitude cloud-resolving simulations considered in \citealt{rasmussen_changes_2017}, and 6--14\%/K in the tropics in the super-parametrized GCM output of \citealt{singh_increasing_2017}). In present-day profiles, variation in upper tropospheric temperature is small relate to that at the  surface, so that warmer surface temperatures are associated with steeper environmental lapse rates (Supplementary Figure S12), enhancing CAPE temperature sensitivity. In future conditions, upper tropospheric warming should follow or even exceed surface warming. However, we expect that the assessment of distributional characteristics of CAPE that provides a useful diagnostic of model performance here may also help in attributing causes of future changes in CAPE.

\medskip

\textit{Acknowledgements.} 
 The authors thank Zhihong Tan for valuable discussion and insight. This work is supported by the Center for Robust Decision-making on Climate and Energy Policy (RDCEP), which is funded by the NSF Decision Making Under Uncertainty program award \#0951576.  This work was completed in part with resources provided by the University of Chicago Research Computing Center. The authors would like to thank the National Center for Atmospheric Research (NCAR) for providing the WRF dataset that made this article possible.

%%%%%%%%%%%%%%%%%%%%%%%%%%%%%%%%%%%%%%%%%%%%%%%%%%%%%%%%%%%%%%%%%%%%%
% REFERENCES
%%%%%%%%%%%%%%%%%%%%%%%%%%%%%%%%%%%%%%%%%%%%%%%%%%%%%%%%%%%%%%%%%%%%%
% Make your BibTeX bibliography by using these commands:
\bibliographystyle{ametsoc2014}
\bibliography{references}

%%%%%%%%%%%%%%%%%%%%%%%%%%%%%%%%%%%%%%%%%%%%%%%%%%%%%%%%%%%%%%%%%%%%%
% TABLES
%%%%%%%%%%%%%%%%%%%%%%%%%%%%%%%%%%%%%%%%%%%%%%%%%%%%%%%%%%%%%%%%%%%%%
%% Enter tables at the end of the document, before figures.
%%
%
%\begin{table}[t]
%\caption{This is a sample table caption and table layout.  Enter as many tables as
%  necessary at the end of your manuscript. Table from Lorenz (1963).}\label{t1}
%\begin{center}
%\begin{tabular}{ccccrrcrc}
%\hline\hline
%$N$ & $X$ & $Y$ & $Z$\\
%\hline
% 0000 & 0000 & 0010 & 0000 \\
% 0005 & 0004 & 0012 & 0000 \\
% 0010 & 0009 & 0020 & 0000 \\
% 0015 & 0016 & 0036 & 0002 \\
% 0020 & 0030 & 0066 & 0007 \\
% 0025 & 0054 & 0115 & 0024 \\
%\hline
%\end{tabular}
%\end{center}
%\end{table}

%%%%%%%%%%%%%%%%%%%%%%%%%%%%%%%%%%%%%%%%%%%%%%%%%%%%%%%%%%%%%%%%%%%%%
% FIGURES
%%%%%%%%%%%%%%%%%%%%%%%%%%%%%%%%%%%%%%%%%%%%%%%%%%%%%%%%%%%%%%%%%%%%%
%% Enter figures at the end of the document, after tables.
%%
%
%\begin{figure}[t]
%  \noindent\includegraphics[width=19pc,angle=0]{figure01.pdf}\\
%  \caption{Enter the caption for your figure here.  Repeat as
%  necessary for each of your figures. Figure from \protect\cite{Knutti2008}.}\label{f1}
%\end{figure}

\end{document}